\documentclass[a4paper,10pt]{article}

\usepackage[english]{babel}
\usepackage{amsmath}
\usepackage{hyperref}
\usepackage{graphics}
\usepackage{bm}
\usepackage{color}
\usepackage{relsize,amsmath,dsfont,mathrsfs,verbatim,upgreek,braket,amscd}
\usepackage{amsfonts}
\usepackage{amssymb}
\usepackage{graphicx}
\usepackage{tabularx,authblk}
\usepackage[margin=1.24in]{geometry}

\newcommand{\R}{\mathbb{R}}
\newcommand{\C}{\mathbb{C}}

\newcommand{\sys}{\mathcal{S}}
\newcommand{\h}{\mathcal{H}}
\newcommand{\hsys}{\h_{\sys}}

\newcommand{\hbarra}{\overline{\h}}
\newcommand{\YUPPI}{\mathlarger{\uppi}}
\providecommand{\bgreek}[1]{\mbox{\boldmath$#1$}}
\newcommand{\vett}[1]{\mathbf{#1}}
\newcommand{\Daw}{\mathcal{D}\left( \frac{1}{2\sqrt{a}} \right)}

\title{Single-Photon Observables and Preparation Uncertainty Relations}

\author[1]{G. Guarnieri}
\author[1]{M. Motta}
\author[1]{L. Lanz \thanks{ludovico.lanz@unimi.it}}

\affil[1]{Dipartimento di Fisica, Universit\` a degli Studi di Milano, Via Celoria 16,
I-20133, Milano, Italy.}

\begin{document}
\maketitle

\begin{abstract}
We propose a procedure for defining all single-photon observables in terms of Positive-Operator Valued Measures (POVMs), in particular spin and position. We identify the suppression of $0$-helicity photon states as a projection from an extended Hilbert space onto the physical single-photon Hilbert space.
We show that all single-photon observables are in general described by POVMs, obtained by applying this projection to opportune Projection-Valued Measures (PVMs), defined on the extended Hilbert space.
The POVMs associated to momentum and helicity reduce to PVMs, unlike those associated to position and spin, this fact
reflecting the intrinsic unsharpness of these observables.
We finally extensively study the preparation uncertainty relations for position and momentum
and the probability distribution of spin, exploring single photon Gaussian states for several
choices of spin and polarization.
\end{abstract}

\section{Introduction}

The investigation of single-photon properties has experienced an increasing interest over 
the last years \cite{tsang, Sciarrino, Birula, Chiao, Tomita, Bomba}.
One of the reasons has to be sought in the escalating request, in many quantum information 
and cryptography protocols \cite{Chen,Grier}, for highly accurate manipulations of spatial 
and polarization single-photon properties.

The most appropriate description of single-photon observables is the subject of an ongoing 
debate. Photon position, in particular, has been considered a controversial concept since 
T.D. Newton and E.P. Wigner first stated \cite{newton} that no position operator can be 
defined in the usual sense for particles with mass $m=0$. Later, A.S. Wightman 
\cite{wightman} extended the search for a notion of photon localization based on a 
Projection-Valued Measure (PVM), again obtaining a negative result.
This evidence suggested K. Kraus to define the single-photon position observable as a Positive-
Operator Valued Measure (POVM) \cite{KrausPos}. After the appearance of K. Kraus' seminal work, 
several authors proposed alternative definitions of the photon position POVM relying on the 
theory of quantum estimate \cite{Holevo2,Holevo3} or on explicit models describing 
actual measurements performed in photocounting experiments \cite{mandel,tsang,HawtonPOVM}.

Despite such remarkable achievements, the most appropriate single-photon description remains 
controversial, since the above approaches were conceived to solve the specific problem of photon 
localization, and do not appear amenable of an immediate generalization to all other 
single-photon observables. In particular, the photon spin is another notoriously delicate 
topic\cite{sp1,sp2,sp4}, often ignored in favor of the more familiar notions of helicity and 
polarization. A renewed interest, due to recent experimental developments especially concerning 
quantum cryptography protocols \cite{Sciarrino}, for the spin of single photons nevertheless 
calls for an appropriate and manageable description of such observable, in a common picture with 
that of position.

In the present work, we generalize K. Kraus' construction of the single-photon position observable, 
given within G. Ludwig's axiomatic formulation of quantum mechanics \cite{Ludwig}, to a 
formalism in which \textit{all} the fundamental single-photon observables are given in a unified 
way in terms of POVMs \cite{Ludwig,HolevoPOVM}.
Following the construction of the single-photon Hilbert space given by K. Kraus and 
H. Moses \cite{KrausPos,Moses}, based on the representation theory of the Poincar\'e group 
for mass $m=0$ and spin $s=1$ particles, given by E. P. Wigner \cite{Wigner}, we interpret the 
notorious suppression of $0$-helicity photon states as a projection from an extended Hilbert 
space onto the single-photon Hilbert space.

Extending K. Kraus' construction of the single-photon position observable, we show that all 
single-photon observables are described by POVMs obtained by applying this projection to PVMs 
defined on the extended Hilbert space and mutuated from the well-established quantum description 
of relativistic massive particles. 

We provide explicit expressions for the POVMs describing the joint measurement of spin and 
momentum, and of spin and position. The results show that momentum and helicity are described 
by PVMs, while spin and position by POVMs. Such difference naturally reflects the 
well-known\cite{HolevoPOVM,Lahti,Breuer2002,Busch2010} circumstance that POVMs describe \emph{unsharp} 
observables reflecting either practical limits in the precision of measurements (in which case 
POVMs typically correspond to coarse-grained version of PVMs) or inherent difficulties in 
realizing a preparation in which the value of an observable is perfectly defined \cite{Busch1989,Busch1996,Busch2010}.
In particular, the intrinsic unsharpness of position and spin results from the coupling between 
momentum and spin introduced by the suppression of $0$-helicity states, a specific consequence 
of the mass $m=0$ and spin $s=1$ of the photon.

We finally apply this formalism to assess the increase of the statistical character of 
single-photon observables naturally brought along by the intrinsic \textit{unsharpness} of POVMs \cite{Ludwig,Breuer2002}.
For this purpose we investigate preparation uncertainty relations for position and momentum, as well 
as the spin probability distribution.
These quantities are analytically calculated for a broad class of physically meaningful single-photon states, 
namely Gaussian states with definite polarization and projections of Gaussian states with definite 
spin. The reasons behind the choice of Gaussian states range from their great theoretical and 
experimental relevance to the fact that, in the non-relativistic context, they saturate the 
notorious inequality $ \Delta X_j \Delta P_j \geq \frac{\hbar}{2}$, identifying themselves as 
the most suitable candidate to investigate the increment of the statistical character of quantum
theory brought into stage by the POVMs.

Our results show that the emergence of POVMs systematically increases the randomness of the 
outcomes\cite{Massar}. In particular, the inequality $ \Delta X_j \Delta P_j \geq \frac{\hbar}{2}$ 
is saturated only in the limiting case of infinitely sharp states in the momentum space; for any 
finite Gaussian width, we observe instead an increase in the product $ \Delta X_j \Delta P_j $, 
of which we give a fully analytic estimate. We observe a similar increase of randomness in the 
spin probability distribution.

Such increment appears to be a manifestation of the unsharpness of position and spin, and of 
the inherent impossibility of sharply localizing a single photon in a bounded space region 
\cite{deb2,Birula}, and of preparing it with definite spin along a spatial direction 
independent on its momentum \cite{Busch1989}.

The paper is organized as follows: in Section \ref{sec:Formalism} the quantum mechanical description of a single free photon is reviewed; in Section \ref{sec:Uncertainty} the procedure for constructing single-photon observables is delineated and the POVMs and 
probability densities of such observables are explicitly given. Finally, in Section \ref{sec:Results}, a detailed study of the 
preparation uncertainty relations for position and momentum and of the spin probability distributions of Gaussian states is 
presented, and conclusions are drawn in Section \ref{sec:Conclusions}.

\section{Single-Photon States}
\label{sec:Formalism}
 
In the present Section, a detailed review of the single-photon formalism given by K. Kraus in \cite{KrausPos}
will be given. Particular attention will be devoted to the representation theory of the Poincar\'e group 
\cite{Moses,Wigner} and on the introduction of an isomorphism based on the representation of the $SU(2)$ group
for spin $s=1$ particles \cite{Moses,Hamermesh}.
All these elements will provide a framework for the discussion in Section \ref{sec:Uncertainty}, where all 
the single-photon observables observables, including spin, will be defined as POVMs through a unified picture 
generalizing K. Kraus' treatment of the position observable \cite{KrausPos}.
It is worth reminding that the term \emph{photon} does not correspond to a unique notion in literature \cite{Lahti}: 
photons are either treated as spin $s=1$ and mass $m=0$ irreducible representations of the Poincar\'e group \cite{Wigner,McCabe},
or as occupations of electromagnetic field modes. In the present work we will rely on the first approach, which 
naturally brings to the introduction of POVMs.
Wigner's seminal work \cite{Wigner} on the representation of theory of the Poincar\'e group, revealed the existence of a deep 
connection between the symmetries underlying Galilean or special relativity and the measurable quantities of an elementary particle.
 Concretely, the mathematical description of a quantum particle existing in space-time should
reflect its Galilei or Poincar\'e invariance, and consequently the state space of such particle should carry an
irreducible representation, characterized by its spin and mass, of either the Galilei or the Poincar\'e group \cite{
Wigner,Mackey,Moses,Lahti}.
With the purpose of introducing the single-photon formalism, let us start by considering the quantum mechanical 
description of a \textit{non-relativistic} particle with spin $s=1$, which takes place in the Hilbert space:
\begin{equation}
\mathcal{H}_{NR} = \mathcal{L}^2(\R^3) \otimes \C^3 
\end{equation}
Here $\R^3$ is meant to be the momentum space, elements of $\mathcal{H}_{NR}$ are square-integrable $3$-component 
functions:
\begin{equation}
\vett{p} \mapsto 
{\bgreek{\psi}}(\vett{p}) = 
\begin{pmatrix}
\psi^1(\vett{p}) \\
\psi^2(\vett{p}) \\
\psi^3(\vett{p}) \\
\end{pmatrix}
\end{equation}
and $\mathcal{H}_{NR}$ is equipped with the inner product:
\begin{equation}
\braket{{\phi} |{\psi}} = \sum_{j=1}^3 \int d^3p \,  \phi^*_j(\vett{p}) \, \psi_j(\vett{p})
\end{equation}
The roto-translation group admits the familiar \cite{Mackey} unitary representation $(\vett{a},R) \mapsto 
U(\vett{a},R)$ on $\mathcal{H}_{NR}$:
\begin{equation}
\label{spintransf}
\big(U(\vett{a},R) \bgreek{\psi} \big)^j (\vett{p}) = e^{-\frac{i}{\hbar} \, \vett{p} \cdot \vett{a}} \, \sum_{k=1}^3 \left( e^{-\frac{i}{\hbar} \phi \vett{n} \cdot \vett{S} } \right)^j_{\,\,\,k} (\bgreek{\psi})^k(R^{-1}\vett{p})
\end{equation}
where $R$ is the matrix associated to the rotation of an angle $\varphi$ around the axis $\vett{n}$, $\vett{a}$ 
is the vector associated to a spatial translation, and $\vett{S}$ denotes the vector of $3\times 3$ Pauli matrices:
\begin{equation}
S_x = \frac{\hbar}{\sqrt{2}}
\begin{pmatrix}
0 & 1 & 0 \\
1 & 0 & 1 \\
0 & 1 & 0
\end{pmatrix}\,
S_y = \frac{\hbar}{\sqrt{2}}\begin{pmatrix}
 0 & -i &  0 \\
 i &  0 & -i \\
 0 &  i &  0
\end{pmatrix} \,
S_z = \,\hbar\,\begin{pmatrix}
1 & 0 & 0 \\
0 & 0 & 0 \\
0 & 0 & -1
\end{pmatrix}
\end{equation}
It is a well-known fact that a projective irreducible representation of the Galilei group can be obtained by introducing the 
non-relativistic position operator $ \vett{X} = i\hbar\partial_{\vett{p}} $ \cite{Mackey,Levy}.

The photon anyway is a purely relativistic particle and thus the formalism given above cannot suit the purpose of
describing it. In order to accomplish such task, let us observe that the matrices $R$ and $\vett{S}$ are related 
to each other by \cite{footnoteV}
\begin{equation}
\label{su2so3}
R = V^{\dagger}\, e^{-\frac{i}{\hbar}\, \varphi \vett{n} \cdot \vett{S}}\, V
\end{equation} 
where the unitary matrix $V$ reads \cite{Hamermesh}
\begin{equation}
\label{vmatrix}
V = 
\begin{pmatrix} 
\frac{1}{\sqrt{2}} & -\frac{i}{\sqrt{2}} & 0 \\
0 & 0 & -1 \\
-\frac{1}{\sqrt{2}} & -\frac{i}{\sqrt{2}} & 0 \\
\end{pmatrix}
\end{equation}
The key relation \eqref{su2so3} is a peculiar prerogative of the spin $s=1$ case and suggests to introduce the 
following unitary transformation:
\begin{equation}
\bgreek{\psi}(\vett{p}) \mapsto {\bgreek{\psi}}_V (\vett{p}) \equiv V^{\dagger} {\bgreek{\psi}}(\vett{p})
\end{equation}
on $\mathcal{H}_{NR}$. It is immediately noticed that wavefunctions ${\bgreek{\psi}}_V (\vett{p}) $ in 
$V \mathcal{H}_{NR}$ transform as a vector field under roto-translations $U_V(\vett{a},R) = V^{\dagger} U(\vett{a},R) V $
\begin{equation}
\label{unirep1}
\big(U_V(\vett{a},R) \bgreek{\psi}_V \big)^j (\vett{p})= e^{-\frac{i}{\hbar} \, \vett{p} \cdot \vett{a}} \, 
\sum_{k=1}^3 R^j_{\,\,\,k} \, \left(\bgreek{\psi}\right)^k_V (R^{-1}\vett{p}\bigr)
\end{equation}
It is worth of notice that the difference between \eqref{unirep1} and \eqref{spintransf} consists in the 
replacements $ \bgreek{\psi} \mapsto \bgreek{\psi}_V $, $ U \mapsto U_V $ and correspondingly $ e^{-\frac{i}{\hbar} 
\phi \vett{n} \cdot \vett{S} } \mapsto R $.
The simple change of basis operated by the matrix $V$ indicates to generalize the formalism outlined above by means 
of the following procedure:
\begin{itemize}
\item endowing wavefunctions in $V \mathcal{H}_{NR}$ with a fourth component, $\psi_V^0(p)$, this way promoting 
them to four-component functions:
\begin{equation}
\label{4compwf}
p \mapsto 
{\psi_V}(p) = 
\begin{pmatrix}
\psi_V^0(p) \\
\psi_V^1(p) \\
\psi_V^2(p) \\
\psi_V^3(p) \\
\end{pmatrix}
\end{equation}
of the four-momentum $p = (p^0,\vett{p})$ satisfying the mass-shell condition $p^\mu g_{\mu \nu} p^\nu = 0$ where 
$g_{\mu\nu} = \mbox{diag}(1,-1,-1,-1)$ and Einstein's convention is assumed. The functions \eqref{4compwf} lie in 
the vector space:
\begin{equation}
\h = \mathcal{L}^2 \left(\R^3 , \frac{d^3 p}{p^0} \right) \otimes \C^4
\end{equation}
\item equipping $\h$ with the following sesquilinear form:
\begin{equation}
\label{scalprod}
\braket{\phi_V |\psi_V} = -\int \frac{d^3 p}{p^0} \, (\phi_V^{\mu})^*(p) \, g_{\mu\nu} \, \psi_V^{\nu}(p)
\end{equation}
in which the Poincar\'e invariant measure $\frac{d^3 p}{p^0}$ replaces the non-relativistic roto-translationally 
invariant Lebesgue measure $d^3 p$.
\item introducing the following linear representation of the Poincar\'e group \cite{Wigner} by isomorphisms on $\h$:
\begin{equation}
\label{unirep2}
\big(U_V(a,\Lambda) \psi_V \big)^{\mu}(p) = e^{\frac{i}{\hbar} a_\tau p^\tau} \Lambda^\mu_{\,\,\, \nu} \, \psi_V^\nu(\Lambda^{-1} p)
\end{equation}
which is a direct generalization of \eqref{unirep1} with $ a $ is a four-vector and $ \Lambda $ is a Lorentz matrix; 
in particular, wavefunctions $\psi_V(p)$ transform as a four-vector under Poincar\'e transformations.
\end{itemize} 

The non-positivity of the scalar product \eqref{scalprod} in the \textit{pseudo-Hilbert space} $\h$, \cite{footnote} 
which is a direct consequence of the non-positivity of the Minkowski inner product $g_{\mu\nu}$, prevents the possibility
of giving a probabilistic interpretation to this formalism. 
Nevertheless, such obstacle can be overcome in the special case of massless particles such as photons. 

In fact, the massless condition naturally identifies a subspace $\sys \subset \h$
\begin{equation}
\label{subsp1}
\sys = \{ \psi_V(p): \, p^\mu g_{\mu \nu} \psi_V^\nu(p) = 0 \,\, \mbox{for almost all} \, p \}
\end{equation}
which is invariant under Poincar\'e transformations, since:
\begin{equation}
\label{poincinvariance}
p^\mu g_{\mu\nu} \left( U_V(a,\Lambda) \psi_V\right)^{\nu} (p) = 
p^\mu g_{\mu\nu} \Lambda^\nu_{\,\,\sigma} \psi_V^{\nu} (\Lambda^{-1}p) = 
(\Lambda^{-1} p)^\mu g_{\mu\nu} \psi_V^{\nu} (\Lambda^{-1}p) = 0
\end{equation}
whence $\psi_V(p) \in \sys \iff U_V(a,\Lambda) \psi_V(p) \in \sys$ for all Poincar\'e transformations. Moreover, the
restriction of \eqref{scalprod} results to be non-negative. To prove this result, it is convenient to expand the 
spatial part of a function $\psi_V(p) \in \sys$ on the \textit{intrinsic frame} $\lbrace \tilde{\vett{e}}_i(\vett{p})
\rbrace_{i=1}^3$ given by:
\begin{equation}
\label{frame}
\begin{split}
\tilde{\vett{e}}_1(p)=\frac{\vett{p} \times (\vett{m} \times \vett{p})}{|\vett{p}| |\vett{m} \times \vett{p}|}
\quad
\tilde{\vett{e}}_2(p)=\frac{\vett{m} \times \vett{p}}{|\vett{m} \times \vett{p}|}
\quad
\tilde{\vett{e}}_3(p)=\frac{\vett{p}}{|\vett{p}|}
\end{split}
\end{equation}
where $\vett{m}$ is an arbitrary unit vector \cite{mandel,riemann_silb} independent on $\vett{p}$. It is worth
of notice that the first elements of the intrinsic frame \eqref{frame} are related to the circular polarization 
vectors $\vett{\tilde{e}}_\pm(\vett{p})$, solution of the eigenvalue equation $\vett{p} \times \tilde{\vett{e}}_
\pm(p) = \mp i |\vett{p}| \tilde{\vett{e}}_\pm(p)$, by the following relation:
\begin{equation}
\tilde{\vett{e}}_\pm(p) = \frac{\tilde{\vett{e}}_1(p) \mp i \tilde{\vett{e}}_2(p)}{\sqrt{2}}
\end{equation}
On this basis, functions in $\sys$ are expressed as:
\begin{equation}\label{psitransv}
\psi_{V}^{\mu}(p) = \begin{pmatrix}
\tilde{\psi}_V^0(p)        \\
\tilde{\bgreek{\psi}}_V(p)
\end{pmatrix}
\end{equation}
where:
\begin{equation}
\tilde{\bgreek{\psi}}_V(p) = \sum_{i=1}^3 \tilde{\psi}_V^i(p) \tilde{\vett{e}}_i(p)
\end{equation}
The condition $\psi_V(p) \in \sys$ translates into $\tilde{\psi}_V^0(p)=\tilde{\psi}_V^3(p)$, and the restriction of
the sesquilinear form \eqref{scalprod} onto $\sys$ reads:
\begin{equation}
\label{scalprodS}
\braket{\phi_V|\psi_V} = \int \frac{d^3 p}{p^0} \left[ (\tilde{\phi}^1_V)^*(p) \, \tilde{\psi}_V^1(p)+(\tilde{\phi}^2_V)^*(p) \, \tilde{\psi}_V^2(p)\right]
\end{equation}
Equation \eqref{scalprodS} is manifestly non-negative, and does not involve the components $\tilde{\psi}_V^0(p),
\tilde{\psi}_V^3(p)$. Following \cite{Moses,KrausPos} we interpret these components as irrelevant degrees of 
freedom, this fact being precursive of the gauge symmetry of the electromagnetic theory\cite{Gupta,Bleuler}.
This interpretation is confirmed if the action of Poincar\'e transformations on functions in $\sys$ is taken in
consideration.
In fact, it is immediate to show that that $\left(U_V(a,\Lambda) \psi_V \right)^{\mu}(p)$ has components
$\left( U_V(a,\Lambda) \tilde{\psi}_V \right)^{1,2}(p)$ which depend only on $\tilde{\psi}_V^{1,2}(p)$ through 
the relation:
\begin{equation}\label{eq:eq1}
\left( U_V(a,\Lambda) \tilde{\psi}_V \right)^i(p) =
\sum_{j=1}^2 \vett{\tilde{e}}_i(p) \cdot \bgreek{\Lambda} \vett{\tilde{e}}_j(\Lambda^{-1} p) \,\,(\tilde{\psi}_V)^j(\Lambda^{-1} p)
\end{equation}
where $\bgreek{\Lambda}$ indicates the spatial part of $\Lambda^\mu_{\,\,\nu}$.
These arguments lead immediately to the construction of the single-photon Hilbert space: equation \eqref{scalprodS}
defines a seminorm on $\sys$, and we can construct a normed space out of $\sys$ taking the quotient $\sys / \sim$ of 
$ \sys $ by the equivalence relation:
\begin{equation}
\phi_V \sim \psi_V \quad \Leftrightarrow \|\phi_V-\psi_V \|=0
\end{equation}
where $ \| \cdot \| $ denotes the seminorm induced by \eqref{scalprod} on $\sys$ \cite{KrausPos}. $\sys / \sim$ is equipped with
the scalar product \eqref{scalprod}, which is now manifestly positive-definite, and carries an irreducible representation of the 
Poincar\'e group defined by \eqref{eq:eq1}.
Each wavefunction in $\hsys$ is an equivalence class of functions ins $\sys$, parametrized by a pair:
\begin{equation}
\begin{pmatrix}
\tilde{\psi}_V^1(p) \\
\tilde{\psi}_V^2(p) \\
\end{pmatrix}
\end{equation}
of complex-valued square-integrable functions, thus making $\sys / \sim$ isomorphic to $ \mathcal{L}^2\left( \R^3,\frac{d^3 
\vett{p}}{|\vett{p}|}\right)\otimes \C^2$ \cite{Segal,KrausPos}.
Each and every equivalence class in $\sys / \sim$ has a representative with the following \emph{transversal form}:
\begin{equation}
\psi_V(p) =
\begin{pmatrix}
0 \\
\tilde{\bgreek{\psi}}_V(p) \\
\end{pmatrix}
\quad\quad \vett{p} \cdot \tilde{\bgreek{\psi}}_V(p) = 0 \to 
\tilde{\bgreek{\psi}}_V(p) = \sum_{i=1}^2 \tilde{\psi}_V^i(p) \tilde{\vett{e}}_i(\vett{p})
\end{equation}
All other elements of the equivalence class are related to the transversal representative by the addition of an unphysical 
component $\tilde{\psi}_V^3(p) (1,0,0,1)^T$. This fact shows that the single-photon Hlbert space is also isomorphic to the
space of square-integrable transverse wavefunctions:
\begin{equation}
\label{thespace}
\hsys = \left\{ \tilde{\bgreek{\psi}}_V(p) : \tilde{\bgreek{\psi}}_V(p) =  
\sum_{i=1}^2 \tilde{\psi}_V^i(p) \tilde{\vett{e}}_i(\vett{p}), \tilde{\psi}_V^i(p) \in 
\mathcal{L}^2\left( \R^3,\frac{d^3\vett{p}}{|\vett{p}|}\right) 
\right\}
\end{equation}
retrieving the result given by K. Kraus in equations (2),(3) of \cite{KrausPos}.

The construction outlined above has lead in a very natural way to the introduction of the single-photon state space 
$\hsys$ by just considering the conditions of mass $m=0$ and spin $ s=1 $, and requiring that $\hsys$ carries an
irreducible representation of the Poincar\'e group. The concrete realization of the single-photon Hilbert space $\hsys$ 
has the advantage of providing a one-to-one correspondence between states and transverse vector functions.

The physically relevant photon states with linear and circular polarization are elements of $\hsys$ of the form:
\begin{equation}
\label{lin_pol1}
\begin{pmatrix}
\tilde{\psi}_V^1(p) \\
\tilde{\psi}_V^2(p) \\
\end{pmatrix} = 
\begin{pmatrix}
\alpha \\
\beta \\
\end{pmatrix} \, \tilde{\psi}_V(p) \quad \alpha,\beta \in \R
\end{equation}
and:
\begin{equation}
\label{lin_pol2}
\begin{pmatrix}
\tilde{\psi}_V^1(p) \\
\tilde{\psi}_V^2(p) \\
\end{pmatrix} = 
\begin{pmatrix}
\phantom{\pm} \frac{1}{\sqrt{2}} \\
\pm \frac{i}{\sqrt{2}} \\
\end{pmatrix} \, \tilde{\psi}_V(p)
\end{equation}
Finally, one might wish to relate such construction of the single-photon state space to the familiar one based on 
the requirement of helicity $\pm 1$ \cite{weinberg}. The equivalence of the two approaches is readily proved by 
recalling the isomorphism $V$, introduced to identify the spin as generator of rotations in \eqref{su2so3}. In fact, 
it can be explicitly shown that the images of the intrisic frame vectors \eqref{frame} under the action of the 
matrix $V$ are closely related to the eigenstates of the helicity operator:
\begin{equation}
\epsilon = \frac{1}{\hbar}\left(\vett{S} \cdot \frac{\vett{p}}{|\vett{p}|}\right) = 
\frac{1}{|\vett{p}|}
\begin{pmatrix}
p_z                                      & \frac{p_x-ip_y}{\sqrt{2}}   & 0 \\
\frac{p_x+ip_y}{\sqrt{2}} & 0                                          & \frac{p_x-ip_y}{\sqrt{2}} \\
0                                          & \frac{p_x+ip_y}{\sqrt{2}} & -p_z \\
\end{pmatrix}
\end{equation}
In particular $ V \tilde{\vett{e}}_3(p) $ is the eigenstate relative to the eigenvalue $ 0 $, while 
$V \tilde{\vett{e}}_{\pm}(\vett{p})$ are the eigenstates relative to eigenvalues $\pm 1$ respectively.

\section{Single-Photon Observables}
\label{sec:Uncertainty}

\subsection{The Extended Hilbert Space}

In the present section single-photon observables will be introduced as POVMs, generalizing the treatment
of position given by K. Kraus in \cite{KrausPos}. With such purpose, the single-photon Hilbert space 
$\hsys$ introduced in the previous section \eqref{sec:Formalism} has to be regarded as a subspace of the 
extended Hilbert space $ \mathcal{H}_A = \mathcal{L}^2(\R^3,\frac{d^3p}{|\vett{p}|}) \otimes \C^3$ whose elements 
have the form:
\begin{equation}
\label{hbarradiA}
\mathbf{f}_V(\vett{p}) = \sum_{i=1}^3 \tilde{\psi}_V^i(\vett{p}) \tilde{\vett{e}}_i(\vett{p})
\end{equation}
and thus differ from elements \eqref{thespace} of $\hsys$ by the addition of a longitudinal component 
$\tilde{\psi}_V^3(\vett{p}) \, \tilde{e}_3(\vett{p})$. 
In equation \eqref{hbarradiA}, and in the remainder of the present work, we express states as functions of $\vett{p}$ rather than $p$ with harmless abuse of notation.

\begin{figure}[htbp!]
\label{fig:spaces}
\begin{center}
\includegraphics[scale=0.19]{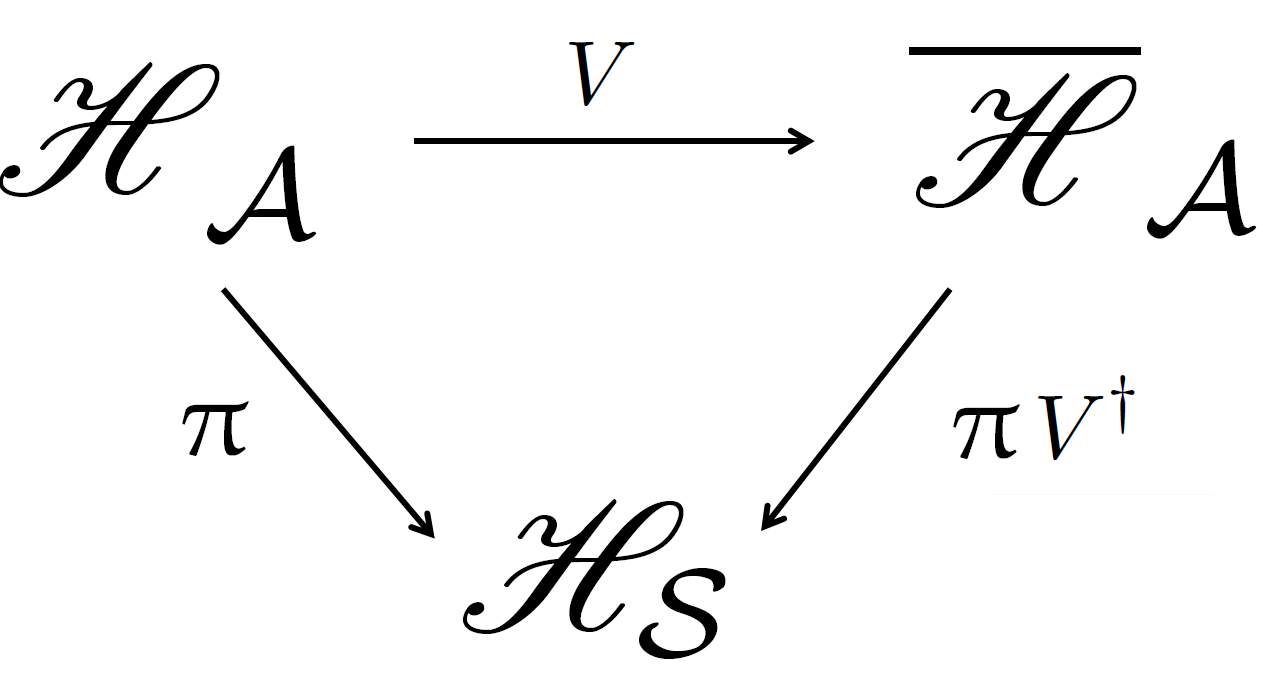}
\end{center}
\caption{Diagram illustrating the action of the isomorphism $ V $ and projection $ \YUPPI $. Wavefunctions 
in $ \h_{\mathcal{S}} $ are represented by couples of complex-valued functions. 
States $ \vett{f}_V$ in $\h_{\mathcal{A}}$ transform according to \eqref{unirep1} under roto-translations,
while states $\vett{f}$ in $ \hbarra_{\mathcal{A}} $ according to \eqref{eq:rototr_hbarraA}.}
\end{figure}

$\hsys$ is obtained from $\mathcal{H}_A$ by means of the projection operator:
\begin{equation}
\label{GB}
\YUPPI : \mathcal{H}_A \to \hsys , \quad (\YUPPI \mathbf{f}_V)^j(\vett{p}) = \sum_k
\YUPPI^j_{\,\,k}(\vett{p}) f_V^k(\vett{p}) \qquad \forall \, \vett{p} \in \R^3
\end{equation}
with
\begin{equation}
\label{GB2}
\YUPPI^j_{\,\,k}(\vett{p}) = \delta^j_{\,\,k} - \frac{p^j p_k}{|\vett{p}|^2} \qquad \forall \, \vett{p} \in \R^3
\end{equation}
The projector \eqref{GB2} eliminates the longitudinal component of the triple $\{ \tilde{\psi}_V^i
(\vett{p}) \}_{i=1}^3$ and can thus be interpreted as an analogue of the Helmholtz projection used
for decomposing the electric and magnetic field of classical electrodynamics into a longitudinal 
and a transverse component.
Under roto-translations, states in $\h_A$ must transform as vector fields:
\begin{equation}
\label{eq:rototr_hA}
U_V(\vett{a},R) \vett{f}_V(\vett{p}) = 
e^{- \frac{i}{\hbar} \vett{a} \cdot \vett{p} } 
R \, \vett{f}_V(R^{-1} \vett{p})
\end{equation}
in order to retrieve \eqref{eq:eq1}.
The introduction of the spin and helicity observables, nevertheless, 
requires a transformation law under roto-translations involving the 
vector of Pauli matrices, as in \eqref{spintransf}, instead of the 
rotation matrix, as in \eqref{unirep1}.
The comparison between equations \eqref{spintransf} and \eqref{unirep1} 
suggests that, to respond to this need, the isomorphism V must be used.
Throughout the remainder of the present work, we will denote using the 
symbol $\hbarra_A =V\h_A$ the isomorphic image of $\h_A$ through the 
isomorphism $V$ and make use of the fact that $\hsys$ corresponds to the image of $\hbarra_A$ through the action of the operator $\YUPPI V^\dagger$, see Figure 
\ref{fig:spaces}.

The change of basis operated by the matrix $V$ not only elucidates the 
equivalence between the conditions of transversality and non-zero helicity 
but, as in the case of non-relativistic particles, leads to an irreducible 
representation of the roto-translation group that involves the vector 
$\vett{S}$ of spin matrices. 
In fact, for a state $\mathbf{f}(\vett{p})$ in $\hbarra_A$, equations \eqref{unirep2} and \eqref{su2so3} 
imply:
\begin{equation}
\label{eq:rototr_hbarraA}
U(\vett{a},R) \vett{f}(\vett{p}) = 
e^{- \frac{i}{\hbar} \vett{a} \cdot \vett{p} } e^{-i \varphi \vett{n} \cdot \vett{S}} \, \vett{f}(R^{-1} \vett{p})
\end{equation}
Elements of $\hbarra_A$ therefore transform under roto-translations according 
to a formula that closely resembles \eqref{spintransf}, holding in 
non-relativistic context.

\subsection{Observables as POVMs on $ \hsys $}

Having constructed the single-photon state space, and embedded it into a suitable extended Hilbert space,
we will now define single-photon observables.
To begin, let us consider the case of massive relativistic particles with spin $s$, where each physical 
observable $\mathcal{O}$ taking values in a sample space $\Omega \subseteq \R$ equipped with 
a suitable sigma-algebra $\mathcal{G}$ is described in terms of a self-adjoint operator $\hat{O}$ on 
$\mathcal{L}\left( \R^3; \frac{d^3p}{p^0} \right) \otimes \C^{2s+1}$. By making use of the spectral 
theorem \cite{Beltrametti,Moretti}, the probability that $\mathcal{O}$ takes values in a measurable set $\mathcal{M} 
\in \mathcal{G}$ is given by the expectation value of a projector $\hat{E}_{\mathcal{O}}(\mathcal{M})$. 
The function $\mathcal{M} \mapsto \hat{E}_{\mathcal{O}}(\mathcal{M})$ is notoriously a PVM \cite{Beltrametti,Moretti}.

This construction can be generalized to a set $\mathcal{O}_1 \dots \mathcal{O}_n$
of $n$ compatible observables, each taking values in the sample space $\Omega_i \subseteq \R$ equipped
with a suitable sigma-algebra $\mathcal{G}_i$. The compatible observables $\mathcal{O}_1 \dots \mathcal{O}_n$,
in fact, take value in the sample space $\Omega_1 \times \dots \times \Omega_n \subseteq \R^n$, equipped with 
the product sigma-algebra $\mathcal{G}_1 \times \dots \times \mathcal{G}_n$.
They are described by a set $\hat{O}_1 \dots \hat{O}_n$ of $n$ self-adjoint commuting operators on $\mathcal{L}
\left( \R^3; \frac{d^3p}{p^0} \right) \otimes \C^{2s+1}$, and also in such situation, the spectral theorem allows 
to express the probability that $\mathcal{O}_1 \dots \mathcal{O}_n$ take values in a measurable set $\mathcal{M} 
\in \mathcal{G}_1 \times \dots \times \mathcal{G}_n$ as the expectation value of a projector 
$\hat{E}_{\mathcal{O}_1 \dots \mathcal{O}_n}(\mathcal{M})$. The map $\mathcal{M} \mapsto \hat{E}_{\mathcal{O}_1 \dots \mathcal{O}_n}(\mathcal{M})$ is referred to
as the joint PVM of the compatible observables $\mathcal{O}_1 \dots \mathcal{O}_n$, and the individual 
PVMs associated to the various observables $\mathcal{O}_i$ are readily obtained as marginals of the joint
PVM through the operation:
\begin{equation}
\hat{E}_{\mathcal{O}_i}(\mathcal{M}_i) = \hat{E}_{\mathcal{O}_1 \dots \mathcal{O}_i \dots \mathcal{O}_n}
(\Omega_1 \times \dots \times \mathcal{M}_i \times \dots \times \Omega_n)
\end{equation}

For massive particles, the fundamental observables of spin, momentum and position are described by the self-adjoint
operators:
\begin{align}
\label{gamma_mom_mass}
\big( \hat{\vett{P}} \mathbf{f} \big)_j (p) &= \vett{p} \, f_j(p) \notag \\
\big( \hat{\vett{S}} \mathbf{f} \big)_j (p) &= \sum_k \vett{S}_{jk} f_k(p) \\
\big( \hat{\vett{X}} \mathbf{f} \big)_j(p) &= 
i \hbar \, \frac{\partial  f_j(p)}{\partial\vett{p}} - \frac{i \hbar}{2} \, \frac{\vett{p}}{\left(p^0\right)^2} \, f_j(p)\notag 
\end{align}
where $\hat{\vett{X}}$ is the well-known Newton-Wigner position operator \cite{newton}. These observables 
are notoriously covariant under roto-translations.

Motivated by the seminal work by K. Kraus \cite{Kraus1983,KrausPos}, we observe that the theory outlined 
in section \ref{sec:Formalism} enables to define, for each single-photon observable $\mathcal{O}$, a 
self-adjoint operator $\hat{O}$ on the Hilbert space $\hbarra_{\mathcal{A}}$ with the same structure as in 
the case of massive particles, e.g.
\begin{align}
\label{gamma_mom}
\big( \hat{\vett{P}} \mathbf{f} \big)_j (p) &= \vett{p} \, f_j(p) \notag \\
\big( \hat{\vett{S}} \mathbf{f} \big)_j (p) &= \sum_k \vett{S}_{jk} f_k(p) \\
\big( \hat{\vett{X}} \mathbf{f} \big)_j(p) &= 
i \hbar \, \frac{\partial  f_j(p)}{\partial\vett{p}} - \frac{i \hbar}{2} \, \frac{\vett{p}}{|\vett{p}|^2} \, f_j(p)\notag
\end{align}
the only difference with \eqref{gamma_mom_mass} being that the massless condition implies $p^0 = |\vett{p}|$.
In the light of equation \eqref{eq:rototr_hbarraA} it is evident that the observables \eqref{gamma_mom} share 
the roto-translational covariance of \eqref{gamma_mom_mass}.

Nevertheless, since the physical space of a single photon is $\hsys$, the image of $\hbarra_{\mathcal{A}}$ 
through the operator $\YUPPI V^\dag$, the probability that $\mathcal{O}$ takes values in a Borel set 
$\mathcal{M} \in \mathcal{G}$ is the expectation value of the \emph{positive operator}: 
\begin{equation}
\label{pvm2povm}
\hat{F}_O(\mathcal{M}) = \left( \YUPPI V^{\dagger} \right) \hat{E}_O(\mathcal{M}) \left(V \YUPPI \right) = \hat{\Omega}^{\dagger}_O(\mathcal{M}) \hat{\Omega}_O(\mathcal{M})
\end{equation}
where $ \hat{\Omega}_O(\mathcal{M}) = \hat{E}_O(\mathcal{M}) V \YUPPI $ and the equality holds by virtue of 
the idempotence of the PVM $ \hat{E}_O(\mathcal{M}) $.
The PVM $\mathcal{M} \mapsto \hat{E}_O(\mathcal{M})$
is turned into the POVM $\mathcal{M} \mapsto \hat{F}_O(\mathcal{M})$ \eqref{pvm2povm} on $\hsys$.
In fact, the operators $\hat{F}_O(\mathcal{M})$, referred to as \emph{effect operators} in published literature,
are still positive and bounded by the identity operator, but in general are not projectors. 
The idempotence property characterizing PVMs is recovered if only if the single-photon Hilbert space $\hsys$ 
is invariant under the action of the projectors $\hat{E}_O(\mathcal{M})$.
In the remainder of the present section, we will show that relevant examples of such \emph{projective or 
sharp observables} observables are momentum, polarization and helicity. On the other hand, we will show that 
the relevant position and spin observables are described by POVMs.

In published literature \cite{Lahti,Busch1996,Busch2009,Busch2010}, observables described by POVMs 
are referred to as \emph{unsharp}, since the emergence of POVMs reflects either practical limits in 
the precision of measurements performed on the system (in which case POVMs appear as coarse-grained 
versions of PVMs) \cite{HawtonPOVM,mandel} or the inherent impossibility of realizing a preparation 
in which the value of an observable can be perfectly defined \cite{HolevoPOVM,Lahti,Busch1996}.

In the case of photons, after explicitly constructing the PVMs and POVMs associated to the fundamental 
single-photon observables, we will interpret the need of describing position and spin as POVMs rather
than PVMs as a consequence of the elimination of $0$-helicity states.

In the context of open quantum systems' theory, POVMs are obtained as projections of opportune PVMs defined on larger Hilbert spaces; from a mathematical point of view, this situation is described by the so-called Naimark's dilation theorem\cite{Ludwig,Holevo,Kraus1983,Breuer2002}.
Within the formalism presented in the present work, all this construction emerges in a very natural way by the treatment of the single-photon Hilbert space exposed in Section \ref{sec:Formalism}.

\subsection{Fundamental POVMs and Probability Distributions}

In the remainder of the present section, we apply the general procedure \eqref{pvm2povm} to explicitly show 
fundamental examples of POVMs.

\subsubsection{Joint probability distribution of $S_z$ and $\vett{P}$} 
As first application, we consider the momentum and spin-$z$ observables, which admit the representation 
\eqref{gamma_mom} on $\hbarra_{\mathcal{A}}$, giving rise to the familiar joint PVM:
\begin{equation}
(\mathcal{M},\hbar m_s) \mapsto 
\left(\hat{E}_{P,S_z}(\mathcal{M},\hbar m_s) \bgreek{\psi} \right)_{s'}(\vett{p}) = 
1_{\mathcal{M}}(\vett{p}) \, \delta_{s,s'} \, \psi_s(\vett{p}) 
\end{equation}
where $1_\mathcal{M}(\vett{p})$ is the indicator function of the Borel set $\mathcal{M}$ and $m_s=2-s$ with $s=1,2,3$. 
The corresponding POVM on $\hsys$, obtained by applying \eqref{pvm2povm} to such PVM, reads:
\begin{equation}
(\mathcal{M},\hbar m_s) \mapsto 
\left(\hat{F}_{P,S_z}(\mathcal{M},\hbar m_s) \bgreek{\psi} \right)_{s'}(\vett{p}) =
1_{\mathcal{M}}(\vett{p}) \, \delta_{s,s'} \, \sum_{i=1}^2 \tilde{\psi}_V^i(\vett{p}) \left[ V\tilde{\mathbf{e}}_i(\vett{p}) \right]_s 
\end{equation}
and gives rise to the following joint probability distribution:
\begin{equation}
\label{pp}
\begin{split}
&p(\vett{p} \in \mathcal{M},S_z = \hbar m_s) = \langle \psi | \hat{F}_{P,S_z}(\mathcal{M}, \hbar m_s) | \psi \rangle \\
=&\, \| \hat{\Omega}_{P,S_z}(\mathcal{M},\hbar m_s) \psi \|^2 = \int_{\mathcal{M}} \frac{d^3 p }{|\vett{p}|} \,\, \Bigl| \sum_{i=1}^2 \tilde{\psi}_V^i(\vett{p}) \left[ V\tilde{\mathbf{e}}_i(\vett{p}) \right]_s\Bigr|^2 \\
\end{split}
\end{equation}

\subsubsection{Joint probability distribution of $S_z$ and $\vett{X}$} 
Since the joint eigenfunctions of the Newton-Wigner position and spin-$z$ operators \eqref{gamma_mom} are the following 
elements of $\hbarra_{\mathcal{A}}$:
\begin{equation}
\vett{u}_{\vett{x},s}(\vett{p}) = \sqrt{|\vett{p}|} \, \frac{e^{-\frac{i}{\hbar} \vett{x} \cdot \vett{p}}}{(2\pi\hbar)^{\frac{3}{2}}} \, \mathbf{e}_s
\end{equation}
and the associated PVM is:
\begin{equation}
\label{povm_pos}
(\mathcal{M},\hbar m_s) 
\mapsto
\left( \hat{E}_{X,S_z}(\mathcal{M},\hbar m_s) \bgreek{\psi} \right)_{s'}(\vett{p}') = \int_{\mathcal{M}} d^3x \, 
\left[ \int \frac{d^3 p}{|\vett{p}|} \vett{u}^*_{\vett{x},s}(\vett{p}) \cdot \bgreek{\psi}(\vett{p}) \right] \, 
\left[ \vett{u}_{\vett{x},s}(\vett{p}') \right]_{s'}
\end{equation}
The corresponding POVM on $\hsys$ is obtained applying \eqref{pvm2povm} to such PVM, and yields the following joint 
probability distribution:
\begin{equation}
\label{px}
\begin{split}
p(\vett{X} \in \mathcal{M},S_z = \hbar m_s) &=  \langle \psi | hat{F}_{X,S_z}(\mathcal{M},\hbar m_s) | \psi \rangle \\
=\,\| \hat{\Omega}_{X,S_z}(\mathcal{M},\hbar m_s) \psi \|^2 &= \int_{\mathcal{M}} d^3x \,
\biggl|\,\left[\tilde{\bgreek{\psi}}_V(\vett{x})\right]_s\,\biggr|^2
\end{split}
\end{equation}
where the probability amplitude $\left[\tilde{\bgreek{\psi}}_V(\vett{x})\right]_s$ reads:
\begin{equation}\label{xspinpdf}
\left[\tilde{\bgreek{\psi}}_V(\vett{x})\right]_s =
\int \frac{d^3p}{|\vett{p}|} \, \sqrt{|\vett{p}|} \, \frac{e^{\frac{i}{\hbar} \vett{x}\cdot \vett{p} }}{(2\pi\hbar)^{\frac{3}{2}}} \,
\sum_{i=1}^2 \tilde{\psi}_V^i(\vett{p}) \left[ V\tilde{\mathbf{e}}_i(\vett{p}) \right]_s
\end{equation}
and can be therefore regarded to as the wave-function for the photon in the configuration space \cite{wf_birula,wf_sipe,
wf_adlard,wf_chan}. It can be noticed that, by virtue of \eqref {unirep1}, this amplitude is covariant under 
roto-translations:
\begin{equation}
\tilde{\bgreek{\psi}}_V(\vett{x}) \overset{U(\vett{a},R)}{\longrightarrow}
\tilde{\bgreek{\psi}}'_V(\vett{x}) = e^{-\frac{i}{\hbar} \varphi \vett{n}\cdot \vett{S}}\,\, 
\tilde{\bgreek{\psi}}_V((R^{-1}(\vett{x}-\vett{a}))
\end{equation}
reproducing the expected transformation laws for the probability density of spin and position and, consequently, for
the momenta of such probability distributions. The position POVM obtained as marginal of \eqref{povm_pos} coincides 
with the photon position observable constructed by K. Kraus' in \cite{KrausPos}.

\subsubsection{Probability distribution of Helicity}

On the single-photon Hilbert space $\hsys$, states with definite circular polarization take the form \eqref{lin_pol2}. 
It is easy to verify that such states are eigenstates of the helicity operator $\epsilon$ 
with eigenvalues $\pm 1$. Consequently, the joint POVM for momentum and helicity on $\hsys$ takes the form:
\begin{equation}
\label{PVM_helicity}
(\mathcal{M},\pm 1) \mapsto
\left(\hat{E}_{P,\epsilon}(\mathcal{M},\pm 1) \,\psi \right)(\vett{p}) = 
1_{\mathcal{M}}(\vett{p}) \, 
\left( \frac{\tilde{\psi}^1_V(\vett{p}) \pm i \tilde{\psi}^2_V(\vett{p}) }{\sqrt{2}} \right) \, 
\tilde{\vett{e}}_\pm(\vett{p})
\end{equation}
it is easy to verify that $ \hsys $ is invariant under the action of the operators \eqref{PVM_helicity}, which makes it
a PVM. Clearly, the marginals of \eqref{PVM_helicity} describing the momentum and helicity observables are also PVMs.
In the light of \eqref{PVM_helicity}, the joint probability distribution of momentum and helicity reads:
\begin{equation}
p\left( \vett{p} \in \mathcal{M}, \epsilon = \pm 1 \right) =
\int_{\mathcal{M}} \frac{d^3 p}{|\vett{p}|} \, 
\left| \frac{\tilde{\psi}^1_V(\vett{p}) \pm i \tilde{\psi}^2_V(\vett{p}) }{\sqrt{2}} \right|^2
\end{equation}
from which the probability distribution of helicity is obtained choosing $\mathcal{M} = \R^3$.

\subsubsection{Probability distribution of $S_n$}

The probability distribution of the spin projection $S_n = \vett{S}\cdot\vett{n}$ along an arbitrary spatial direction $\vett{n}$ 
is readily found recalling that the matrix $\vett{S}\cdot\vett{n}$ admits in $\mathbb{C}^3$ the spectral decomposition 
$\vett{S}\cdot\vett{n} = \sum_{m_s} \hbar m_s \ket{\bgreek{\phi}_{\vett{n},m_s}} \bra{\bgreek{\phi}_{\vett{n},m_s}}$. The joint
PVM of $S_n$ and momentum on $\hbarra_{\mathcal{A}}$ reads:
\begin{equation}
(\mathcal{M},\hbar m_s) 
\mapsto
\left(\hat{E}_{P,S_n}(\mathcal{M},\hbar m_s) \bgreek{\psi} \right)_{s'}(\vett{p}) = 
1_{\mathcal{M}}(\vett{p}) \, \left[ \bgreek{\phi}_{\vett{n},m_s} \right]_{s'} \,\, \bgreek{\phi}^*_{\vett{n},m_s} \cdot \bgreek{\psi}(\vett{p}) 
\end{equation}
The corresponding POVM on $\hsys$, obtained by applying \eqref{pvm2povm} to such PVM, leads to the joint
probability distribution:
\begin{equation}
p(\vett{p} \in \mathcal{M},S_n = \hbar m_s) = 
\int_{\mathcal{M}} \frac{d^3 p }{|\vett{p}|} \,\, \Bigl| \sum_{i=1}^2 \tilde{\psi}_V^i(\vett{p}) \,\,
\bgreek{\phi}^*_{\vett{n},m_s} \cdot V\tilde{\mathbf{e}}_i(\vett{p}) \Bigr|^2 \\
\end{equation}
from which $p(S_n = \hbar m_s)$ is readily obtained choosing $\mathcal{M}=\R^3$.

\subsubsection{Uncertainty Relations for Position and Momentum}

In the light of \eqref{px} and \eqref{pp}, it becomes interesting to investigate the preparation uncertainty relations for 
position and momentum observables of a single photon. 
On $\hbarra_{\mathcal{A}}$, where both these observables are defined in terms of self-adjoint operators with usual commutator 
$[X_{NW \, i},P_j] = i \hbar \, \delta_{ij}$, the familiar inequality:
\begin{equation}
\label{heisenberg}
\Delta X_k \Delta P_k \geq \frac{\hbar}{2}
\end{equation}
holds, and is saturated by Gaussian states with definite spin along an arbitrary spatial direction.
Equation \eqref{heisenberg} retains its validity on the physical space $ \hsys $, but, due to the projection $ \YUPPI $, it is no longer saturated 
by the minimum uncertainty states on $ \hbarra_{\mathcal{A}} $.

Mean values and variances will be computed taking the marginals of the joint probability distributions \eqref{px} and \eqref{pp}
over the spin degrees of freedom. 
In the case of position, we are left with the following expression: 
\begin{equation}
\label{pdfpos}
p(\vett{X} \in \mathcal{M}) = \int_{\mathcal{M}} d^3x \, \sum_{s=1}^3 \, \biggl|\,\left[\tilde{\bgreek{\psi}}_V(\vett{x})\right]_s\,\biggr|^2
\end{equation} 
where the configurational wave-function $\tilde{\psi}_V(\vett{x})$ is known to exhibit polynomial\cite{paley_wiener,mandel,deb2} or 
even exponential\cite{Birula} decay, ensuring the existence of the first momenta of \eqref{pdfpos}.

The latter are readily obtained inserting \eqref{xspinpdf} in \eqref{pdfpos}:
\begin{equation}
\begin{split}
\label{nwmomenta}
\langle X_k \rangle &= i \hbar \, \sum_{s=1}^3 \int d^3 p \,\, \frac{\left[\tilde{\bgreek{\psi}}_V(p)\right]_s}{\sqrt{|\vett{p}|}} \, \frac{\partial}{\partial p_k} \left(\frac{\left[\tilde{\bgreek{\psi}}_V(p)\right]_s}{\sqrt{|\vett{p}|}}\right) \\
\langle X^2_k \rangle &= - \hbar^2 \, \sum_{s=1}^3 \int d^3 p \,\, \frac{\left[\tilde{\bgreek{\psi}}_V(p)\right]_s}{\sqrt{|\vett{p}|}}\, \frac{\partial^2}{\partial p_k^2} \left(\frac{\left[\tilde{\bgreek{\psi}}_V(p)\right]_s}{\sqrt{|\vett{p}|}}\right). \\ 
\end{split}
\end{equation}

In the remainder of the work, the quantity $\Delta X_k \Delta P_k$ will be referred to as
uncertainty product.

\subsection{Interpretation}

Equation \eqref{pvm2povm} represents the main contribution of the present
work. It generalizes K. Kraus' construction of the single-photon position 
observable to a formalism in which all the single-photon observables are
defined in a unified way in terms of POVMs.
In the present section, several relevant single-photon observables have 
been considered and the POVMs describing them have been derived, according
to \eqref{pvm2povm}, as projections on $\hsys$ of PVMs defined on 
$\hbarra_{\mathcal{A}}$, whose expression is borrowed form the quantum 
theory of relativistic massive particles.

It is worth of notice that position and spin are described by POVMs, while
momentum and helicity by PVMs, a circumstance which is amenable of a clear 
physical interpretations.

Observables $\mathcal{O}$ represented as PVMs are commonly understood to 
correspond to measurements with perfect accuracy, and are therefore called 
\emph{sharp observables}. 
In fact, the generic projector $\hat{E}_{\mathcal{O}}(\mathcal{M})$ associated
to the event $\mathcal{O} \in \mathcal{M}$ identifies a subspace: 
\begin{equation}
\mathcal{S}_{\mathcal{O}}(\mathcal{M}) = 
\left\{ \ket{\psi}: \hat{E}_{\mathcal{O}}(\mathcal{M}) \ket{\psi} = \ket{\psi} \right\}
\end{equation}
of preparations for which the event $\mathcal{O} \in \mathcal{M}$ has probability
$1$. This property is notoriously not shared by generic POVMs, which are therefore
associated to \emph{unsharp observables} \cite{HolevoPOVM,Busch2009}.
In this sense, we understand that the emergence of POVMs enhances the statistical
character of quantum theory: concretely, the recursion to POVMs typically reflects    
either imperfect measurements or inherent difficulties in realizing a preparation 
in which the event $\mathcal{O} \in \mathcal{M}$ has probability $1$.

In the case of photons, the need of treating position and spin as unsharp
observables reflects the properties of single-photon preparations rather than those
of measurement procedures.

In particular, the suppression of $0$-helicity states determines a coupling
between momentum and spin, expressed by the key role of the helicity, which limits 
the possibility to prepare a photon with definite spin along a fixed spatial 
direction independent on its momentum.
Similarly, since transverse wavefunctions are linear combinations of $\tilde{\vett{e}}_1(\vett{p})$ and $\tilde{\vett{e}}_2(\vett{p})$ and the latter are not analytic functions of $\vett{p}$ \cite{deb2}.
The Paley-Wiener theorem \cite{paley_wiener} poses severe limitations on the possibility of localizing photons in the configuration space \cite{deb2,Birula}.
These inherent difficulties find a direct correspondence in our construction, which highlights the need of treating spin and position as unsharp quantum observables.

Finally, it is worth of notice that Equation \eqref{pvm2povm} can be further generalized to the case of a set of incompatible (non-commuting) observables $\mathcal{O}_1\ldots \mathcal{O}_n$ on $\mathcal{H}_A$.
In this case the probability that $\mathcal{O}_1\ldots \mathcal{O}_n$ take values in a measurable set $\mathcal{M} \in \mathcal{G}_1 \times \dots \times \mathcal{G}_n$ is given in terms of the expectation value of an effect operator $\hat{F'}_{\mathcal{O}_1 \ldots \mathcal{O}_n}(\mathcal{M})$. The family of such operators is referred to as the POVM associated to the unsharp joint measurement of the incompatible observables $\mathcal{O}_1\ldots \mathcal{O}_n$ \cite{HolevoPOVM,Breuer2002}.
The projection onto the physical single-photon space $\hsys$ leads therefore to another POVM which is related to that defined on $\mathcal{H}_A$ by means of the equation:
\begin{equation}\label{eq:povm2povm}
\hat{F}_{\mathcal{O}_1\ldots \mathcal{O}_n}(\mathcal{M}) =  \left( \YUPPI V^{\dagger} \right) \hat{F'}_{\mathcal{O}_1\ldots \mathcal{O}_n}(\mathcal{M}) \left(V \YUPPI \right)
\end{equation}
This expression, which clearly extends \eqref{pvm2povm}, can be used for example to describe the joint measurement of position and momentum observables.

\section{Results and Theoretical Interpretation}
\label{sec:Results}

In the present Section we will consider two classes of physically relevant single-photon states, namely Gaussian states with definite polarization and projections of Gaussian states with definite spin, and extensively investigate: (i) the preparation uncertainty relations $ \Delta X_k \Delta P_k $ and (ii) the probability distribution of spin along a certain direction.

The study of both these properties will highlight and quantitatively estimate the increase of the 
statistical character naturally brought along by the intrinsic \textit{unsharpness} of POVMs 
\cite{HolevoPOVM,Holevo,Ludwig,Massar,Breuer2002}.
The results will show both an increment in the product $\Delta X_k \Delta P_k$
for Gaussian states, which in non-relativistic framework saturate the inequality $\Delta X_k \Delta P_k
\geq \frac{\hbar}{2}$, and the increase of randomness in the probability distribution of spin along
a fixed spatial direction.


\subsection{Gaussian States with Definite Polarization}

Let us consider Gaussian states with definite polarization in $ \h_{\mathcal{S}} \simeq \mathcal{L}^2\left(\R^3,\frac{d^3\vett{p}}{|\vett{p}|}\right)\otimes\C^2 $, i.e. states of the form:
\begin{equation}
\label{psi1}
\begin{pmatrix}
\tilde{\psi}^1_V(\vett{p}) \\
\tilde{\psi}^2_V(\vett{p}) \\
\end{pmatrix}
=
\sqrt{|\vett{p}|} \, \frac{e^{- \frac{|\vett{p}-\vett{p}_0|^2}{8 a p_0^2} }}{\big( 4 \pi a p_0^2 \big)^{\frac{3}{4}}} \, e^{-\frac{i}{\hbar} \vett{p}\cdot\vett{x}_0} \,\,
\begin{pmatrix}
\gamma^1\\
\gamma^2
\end{pmatrix}
\end{equation}
where $\vett{p}_0 = \langle \hat{\vett{P}} \rangle $, $\vett{x}_0 = \langle \hat{\vett{X}} \rangle$,
$\sum_{i=1}^2 |\gamma^i|^2 = 1$ and:
\begin{equation}
a = \frac{(\Delta p)^2}{2 p_0^2}
\end{equation}
is a positive, dimensionless parameter which takes into account the wavefunction's spread in 
momentum space.
In the following, without any loss of generality, the chioce $\vett{p}_0= |\vett{p}_0| e_z$
will be made.

\subsubsection{Preparation Uncertainty Relations for Position and Momentum}

The square norm of states in $ \hsys $ of the form \eqref{psi1} is obtained from a straightforward
calculation and reads:
\begin{equation}
\braket{\tilde{\psi}_V|\tilde{\psi}_V} = \sum_{k=1}^2 |\gamma^k|^2
\end{equation}
so that, if $\sum_{k=1}^2 |\gamma^k|^2 = 1$, the mean values and the variances of momentum 
components are given by:
\begin{equation}
\begin{split}
\langle P^j\rangle &= p_0^j \\
\vspace{0.2in}
\langle (P^j)^2 \rangle &= (p_0^j)^2 + 2a p_0^2 \\
\end{split}
\end{equation}
On the contrary, the calculation of the expectation value and variance of position components is 
complicated by the circumstance that the gradient appearing in \eqref{nwmomenta} acts non-trivially 
on the vectors $\tilde{\vett{e}}_1(\vett{p}), \tilde{\vett{e}}_2(\vett{p})$. Applying formula 
\eqref{nwmomenta} to the states \eqref{psi1}, it is readily found that the mean values of $X_j$ and 
$X_j^2$ are bilinear functions of the coefficients $\gamma^j$, defined by the matrices:
\begin{equation}
\begin{split}
\label{matrices_pol}
\left[X^j \right]_{kl} &= i \hbar \int d^3 p \, g(\vett{p})
\, \tilde{\vett{e}}_k(\vett{p}) \cdot \partial_{p_j} \tilde{\vett{e}}_l(\vett{p}) \\
\left[ (X^j)^2 \right]_{kl} &= - \hbar^2 \, \int d^3 p \, g(\vett{p})
\, \left[ \tilde{\vett{e}}_k(\vett{p}) \cdot \partial^2_{p_jp_j} \tilde{\vett{e}}_l(\vett{p})
-\frac{p_j - p_{0 j}}{4 a p_0^2} \tilde{\vett{e}}_k(\vett{p}) \cdot \partial_{p_j} \tilde{\vett{e}}_l(\vett{p}) \right] + \delta_{kl} \, \frac{\hbar^2}{8 a p_0^2} + (x_0^j)^2 \, \delta_{kl} \\
\end{split}
\end{equation}
with $g(\vett{p}) = \frac{e^{- \frac{|\vett{p}-\vett{p}_0|^2}{4 a p_0^2} }}{\big( 4 \pi a p_0^2 \big)^{\frac{3}{2}}}$.
Performing the integrations \eqref{matrices_pol} using shifted spherical coordinates one finds:
\begin{equation}
\begin{split}
\left[X^j\right]_{kl} &= 0 \quad \quad \forall j = 1,2,3\\
\left[(X^3)^2\right]_{kl} &\equiv \left[(Z)^2\right]_{kl}= \hbar^2 \delta_{kl} \, \frac{1 - 4 a + 4 \sqrt{a} \, (1 + 2 a) \, \Daw}{8 a p_0^2} \\
\left[(X^1)^2\right]_{kl} &\equiv \left[(X)^2\right]_{kl}= \hbar^2 \delta_{kl} \, \frac{1 + 8 a -16 a^{3/2} \Daw }{8 a p_0^2} \\
\left[(X^2)^2\right]_{kl} &\equiv \left[(Y)^2\right]_{kl}= \hbar^2 \delta_{kl} \, \frac{1 + 8 a -16 a^{3/2} \Daw }{8 a p_0^2} \\
\end{split}
\end{equation}
where $ \mathcal{D}(x) $ denotes the Dawson's function (sometimes also referred to as Dawson's Integral) of argument $ x $ \cite{Abra}.

\begin{figure}[htbp!]
\begin{center}
\includegraphics[scale=0.6]{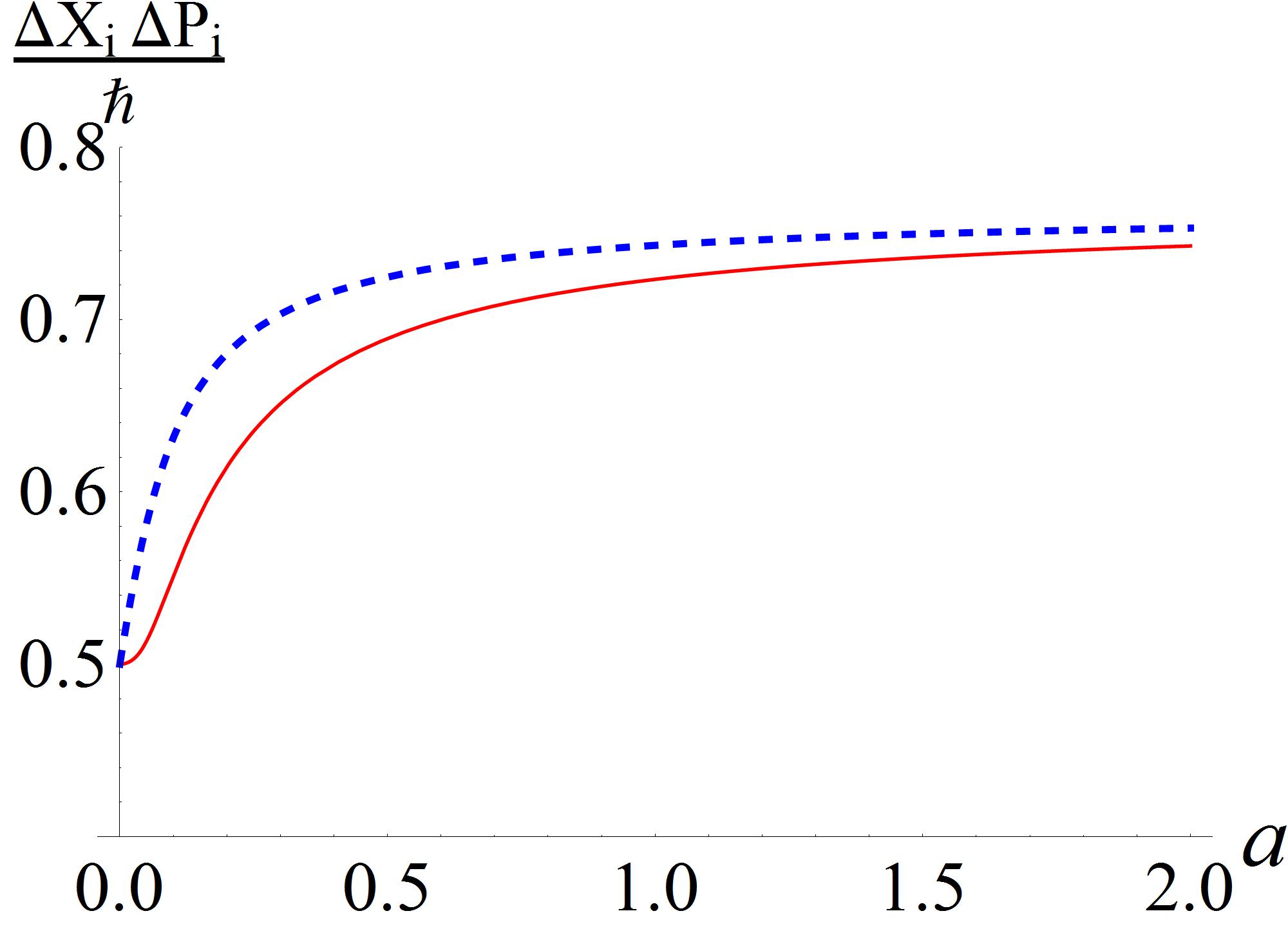}
\end{center}
\caption{(color online): uncertainty products $\Delta X_i \Delta P_i$, in units of $\hbar$, along directions parallel (red solid line) and perpendicular (blue dashed line) to $\vett{p}_0$, for states with definite polarization.}
\label{fig:e1ofp}
\end{figure}

The uncertainty products \emph{both} for $ \begin{pmatrix} \gamma^1\\ \gamma^2 \end{pmatrix}= \begin{pmatrix} 1\\0 \end{pmatrix} $ \emph{and} $ \begin{pmatrix} 0\\1 \end{pmatrix} $, therefore read:
\begin{equation}
\begin{split}
\Delta Z \Delta P_z &= \frac{\hbar}{2} \sqrt{1 - 4 a + 4 \sqrt{a} \, (1 + 2 a) \, \Daw} \\
\Delta X \Delta P_x = \Delta Y \Delta P_y &= \frac{\hbar}{2} \sqrt{1 + 8 a -16 a^{3/2} \Daw } 
\end{split}
\end{equation}
and are illustrated in Figure \ref{fig:e1ofp}.

The results are monotonically increasing functions of the parameter $a$ and remarkably, in the limit $a \to 0$, the minimum value $\Delta X_i \Delta P_i = \frac{\hbar}{2}$ is obtained for every component thus indicating that only for an infinitely sharp wavefunction in the momentum space the familiar Heisenberg limit is retrieved.

In the opposite limit $a \to \infty$, the uncertainty product approach from below an 
asymptotic finite value which is equal to $\sqrt{\frac{7}{12}} \hbar$ for every component.
In table \eqref{table:limiting}, the asymptotic expansions of the uncertainty
products are given.

$ $

\begin{table}[!h]
\caption{Limiting behaviour of the uncertainty product \eqref{heisenberg}, in units of $ \hbar $, for states with definite polarization along directions parallel (z axis) and perpendiculr (x or y axis) to $ \vett{p}_0 $.}
\label{table:limiting}
\begin{center}
\begin{tabular}{c c c}
\vspace{0.001in} \\
\hline\\
axis & small $a$ & large $a$ \tabularnewline
\vspace{0.001in}\\
\hline\\
\vspace{0.001in} \\
z & $\frac{1}{2} + 4a^2 $ & $ \quad\sqrt{\frac{7}{12}} - \frac{1}{5\sqrt{21} a} + \frac{11}{700\sqrt{21} a^2}$ \\
\vspace{0.001in} \\
x (y) & $\frac{1}{2} + 2a - 16 a^2$ & $\quad\sqrt{\frac{7}{12}} - \frac{1}{5\sqrt{21} a} + \frac{1}{175\sqrt{21} a^2}$ \\
\\
\hline\\
\end{tabular}
\label{table:limiting}
\end{center}
\caption{Limiting behaviour of the uncertainty product \eqref{heisenberg}, in units of $ \hbar $, for states with definite polarization along directions parallel (z axis) and perpendiculr (x or y axis) to $ \vett{p}_0 $.}
\end{table}

The increase in the uncertainty product witnessed for $a>0$ is a direct consequence of 
the \textit{unsharpness} of the POVM associated with the position observable, and gives a direct 
example of the circumstance, well known in the quantum theory of measurement \cite{HolevoPOVM,Holevo,Ludwig,Lahti,
Breuer2002}, that POVMs enhance the randomness of measurement outcomes and the statistical character 
of the underlying quantum theory.

\subsubsection{Probability distribution of $S_z$}

The explicit expression of the probability distribution of $S_z$ for Gaussian states of the form \eqref{psi1} will now be given. It reads:
\begin{equation}
\label{pszhel}
p(S_z= \hbar m_s) = \sum_{ij} \, \left(\gamma^i\right)^* \, [\Sigma (m_s)]_{ij} \, \gamma^j
\end{equation}
where $m_s=2-s$ and
\begin{equation}
[\Sigma (m_s)]_{ij} = \int d^3\vett{p} \, \left[ \tilde{\vett{e}}_i(\vett{p}) \right]^*_s
\left[ \tilde{\vett{e}}_j(\vett{p}) \right]_s \,\, 
g(\vett{p})
\end{equation}
Equation \eqref{pszhel} is a quadratic function of $\begin{pmatrix} \gamma^1 \\ \gamma^2 \end{pmatrix}$ and can be expressed in terms of three functions (the first two of which will appear also in the 
forthcoming discussion relative to states with definite spin):
\begin{equation}
\begin{split}
\label{eq:specialfunctions}
u_1(a) &= 1 - 6a + 12 a^{3/2} \, \Daw \\
u_2(a) &= 1 - 2 \sqrt{a} \, \Daw \\
u_3(a) &= \frac{1}{2} \left( \frac{2 \sqrt{a}}{\sqrt{\pi}} e^{-\frac{1}{4a}}+(1-2a) \, 
\mbox{erf}\left( \frac{1}{2\sqrt{a}} \right) \right)
\end{split}
\end{equation}
The matrices $ \Sigma(m_s) $ are given by
\begin{equation}
\label{spin_ddp_hel2}
\begin{split}
& \Sigma (1) = \begin{pmatrix}
\frac{1}{3}+\frac{u_1(a)}{6} & i \,u_3(a) \\
                   -i\, u_3(a) & \frac{1}{3}+\frac{u_1(a)}{6} \\
\end{pmatrix} \\
& {\Sigma (0)} = \begin{pmatrix}
2 a u_2(a) & 0 \\
0          & 2 a u_2(a) \\
\end{pmatrix} \\
& {\Sigma (-1)} = \begin{pmatrix}
\frac{1}{3}+\frac{u_1(a)}{6} & -i u_3(a) \\
                    i u_3(a) & \frac{1}{3}+\frac{u_1(a)}{6} \\
\end{pmatrix}
\end{split}
\end{equation}
from which it can be immediately seen that:

a) the eigenvectors of \eqref{spin_ddp_hel2} are the states $\gamma_\pm$ with definite circular 
polarization

b) $p(S_z=0)$ for $\gamma_+$ and $p(S_z=0)$ for $\gamma_-$ are equal to each other.

c) $p(S_z=\hbar)$ for $\gamma_+$ and $p(S_z=-\hbar)$ for $\gamma_-$ are equal to each other.

\begin{figure}[htbp!]
\center
\includegraphics[scale=0.6]{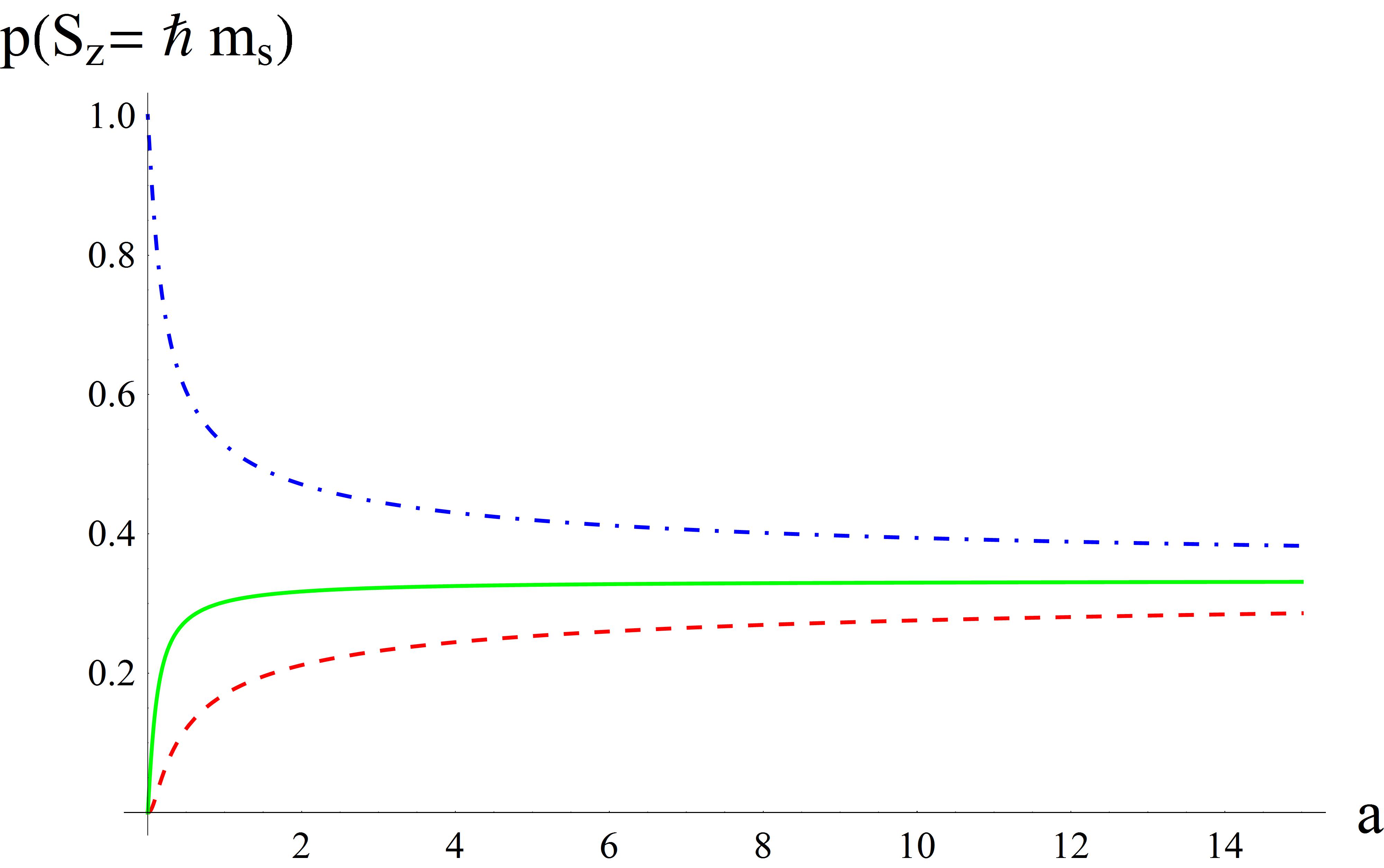}
\caption{Color online: probability distribution of $ S_z $ (red dashed line for $ m_s=1 $, blue dot-dashed line for $ m_s=0 $, green solid line for $ m_s=-1 $) on states with definite polarization $+1$.}
\label{fig:spinddphel}
\end{figure}

Figure \eqref{fig:spinddphel} shows the probability distribution of $S_z$ for a Gaussian state with 
circular polarization $+1$. 

Finally, the asymptotic behavior of the probability distribution of $S_z$ for Gaussian states with definite polarization is listed in Table \eqref{table:szhel}.

\begin{table}[h!]
\begin{center}
\begin{tabular}{c c c c}
\hline \\
$S_z/\hbar$ & polarization & small $a$ & large $a$ \\
\\
\hline\\
 1 & +1 & $2 a^2$     & $\frac{1}{3}-\frac{1}{\sqrt{\pi a}}+\frac{1}{60 a}$ \\ 
 0 & +1 & $2a-4a^2$   & $\frac{1}{3}-\frac{1}{30 a}$ \\
-1 & +1 & $1-2a+2a^2$ & $\frac{1}{3}+\frac{1}{\sqrt{\pi a}}+\frac{1}{60 a}$ \\
 1 & -1 & $1-2a+2a^2$ & $\frac{1}{3}+\frac{1}{\sqrt{\pi a}}+\frac{1}{60 a}$ \\ 
 0 & -1 & $2a-4a^2$   & $\frac{1}{3}-\frac{1}{30 a}$ \\
-1 & -1 & $2 a^2$     & $\frac{1}{3}-\frac{1}{\sqrt{\pi a}}+\frac{1}{60 a}$ \\
\hline
\label{table:szhel}
\end{tabular}
\end{center}
\caption{asymptotic behaviour of the probability distribution of $S_z$ for the choices of $\gamma$ 
reported in figure \ref{fig:spinddphel}.}
\end{table}

\subsection{Gaussian States with Definite Spin}

Gaussian states with definite spin along an arbitrary spatial axis will now be considered, which
are defined in $ \hbarra_{\mathcal{A}} \simeq \mathcal{L}^2\left(\R^3,\frac{d^3\vett{p}}{|\vett{p}|}\right)\otimes\C^3 $ as:
\begin{equation}
\label{spindefinite}
\mathbf{f}(\vett{p}) = \sqrt{|\vett{p}|} \, \frac{e^{- \frac{|\vett{p}-\vett{p}_0|^2}{8 a p0^2} }}{\left( 4 \pi a p_0^2 \right)^{\frac{3}{4}}} \,	\mathbf{h} \quad \quad \mathbf{h} = \begin{pmatrix}
h_1 \\ h_2 \\ h_3
\end{pmatrix} \,, \quad \sum_{i=1}^3 |h_i|^2 = 1
\end{equation}

The corresponding physical states are elements in $ \hsys $ given by:
\begin{equation}
\tilde{{\psi}}_V(\vett{p}) = \frac{\YUPPI V^{\dagger} \mathbf{f}(\vett{p})}{K}
\end{equation}
the quantity $ K^2 = \braket{{\YUPPI} V^{\dagger} \mathbf{f} |{\YUPPI} V^{\dagger} 
\mathbf{f} } = \braket{V {\YUPPI} V^{\dagger} \mathbf{f} |V {\YUPPI} V^{\dagger} \mathbf{f} } $ 
being a normalization constant.

\subsubsection{Preparation Uncertainty Relations for Position and Momentum}

For states in $ \hbarra_{\mathcal{A}} $ of the form \eqref{spindefinite}, denoting for simplicity
$ {\Pi}(\vett{p}) = V {\YUPPI} V^{\dagger} $ we readily find that the square norm $K^2$ and the
first momenta of the position and momentum probability distributions are quadratic functions of
$\mathbf{h}$ defined by the matrices:

\begin{equation}
\begin{split}
\label{matrices}
[K]_{kl} &= \int d^3 p \, g(\vett{p}) 
[\Pi(\vett{p})]_{kl} \\
[P^j]_{kl} &= \int d^3 p \, g(\vett{p})
\, p^j \, [\Pi(\vett{p})]_{kl} \\
[(P^2)^j]_{kl} &= \int d^3 p \, g(\vett{p})
\, (p^j)^2 \, [\Pi(\vett{p})]_{kl} \\
[X^j]_{kl} &= i \hbar\, \int d^3 p \, g(\vett{p})
\, \left[ \frac{\partial}{\partial p_j} \, [\Pi(\vett{p})]_{kl} - \frac{p^j - p_0^j}{4 a p_0^2} \, [\Pi(\vett{p})]_{kl} \right] \\
[(X^2)^j]_{kl} &= - \hbar^2 \int d^3 p \, g(\vett{p})
\, \left[ \frac{\partial^2}{\partial p_j^2} \, [\Pi(\vett{p})]_{kl} - 2 \, \frac{p^j - p_0^j}{4 a p_0^2} \, \frac{\partial}{\partial p_j}[\Pi(\vett{p})]_{kl}\right.\notag\\
&\qquad\qquad\qquad\qquad\qquad\qquad\qquad\qquad+\left.\left[ \left( \frac{p^j - p_0^j}{4 a p_0^2} \right)^2 - \frac{1}{4 a p_0^2} \right] [\Pi(\vett{p})]_{kl} \right]
\end{split}
\end{equation}

of which we below the analytic expression, resulting from a calculations in shifted spherical 
coordinates, which remarkably can be expressed in terms of the functions $u_1(a)$ and $u_2(a)$ 
introduced above in \eqref{eq:specialfunctions}. The normalization reads:
\begin{equation}
K = \left( \begin{array}{ccc}
1 & 0 & 0 \\
0 & 0 & 0 \\
0 & 0 & 1 \\
\end{array} \right) + 2 a \, u_2(a) \, 
\left( \begin{array}{ccc}
-1 & 0 & 0 \\
0 & 2 & 0 \\
0 & 0 & -1 \\
\end{array} \right) ;
\end{equation}
the first and second momenta of the momentum probability distribution are given by:
\begin{align}
\label{meanvaluesP}
&P^1 \equiv P_x = 2p_0 \, a\, u_1(a)\, \left( \begin{array}{ccc}
0 & \frac{1}{\sqrt{2}} & 0 \\
\frac{1}{\sqrt{2}} & 0 & -\frac{1}{\sqrt{2}} \\
0 & -\frac{1}{\sqrt{2}} & 0 \\
\end{array} \right) \\ 
&P^2 \equiv P_y = 2p_0 \, a\,  u_1(a) \, \left( \begin{array}{ccc}
0 & -\frac{i}{\sqrt{2}} & 0 \\
\frac{i}{\sqrt{2}} & 0 & \frac{i}{\sqrt{2}} \\
0 & -\frac{i}{\sqrt{2}} & 0 \\
\end{array} \right) \\ 
&P^3 \equiv P_z = p_0 \left( \begin{array}{ccc}
1 & 0 & 0 \\
0 & 0 & 0 \\
0 & 0 & 1 \\
\end{array} \right) + 2p_0 a \,  u_1(a) \, \left( \begin{array}{ccc}
-1 & 0 & 0 \\
0 & 2 & 0 \\
0 & 0 & -1 \\
\end{array} \right) \\
&(P^1)^2 = (2p_0)^2 \, \frac{a}{2} \, \left( \begin{array}{ccc}
1 & 0 & 0 \\
0 & 0 & 0 \\
0 & 0 & 1 \\
\end{array} \right) + (2p_0)^2 \, a^2 \,  u_1(a)
\left( \begin{array}{ccc}
-2 & 0 & 1 \\
0 & 4 & 0 \\
1 & 0 & - 2 \\
\end{array} \right) \\
&(P^2)^2 = (2p_0)^2 \, \frac{a}{2} \, \left( \begin{array}{ccc}
1 & 0 & 0 \\
0 & 0 & 0 \\
0 & 0 & 1 \\
\end{array} \right) + (2p_0)^2 \, a^2 \,  u_1(a)
\left( \begin{array}{ccc}
-2 & 0 & -1 \\
0 & 4 & 0 \\
-1 & 0 & - 2 \\
\end{array} \right) \\
&(P^3)^2= p_0^2 \left( \begin{array}{ccc}
1 & 0 & 0 \\
0 & 4a & 0 \\
0 & 0 & 1 \\
\end{array} \right) + (2p_0)^2 \, 4 a^2 \,  u_1(a) \, 
\left( \begin{array}{ccc}
1 & 0 & 0 \\
0 & -2 & 0 \\
0 & 0 & 1 \\
\end{array} \right) , \\ \nonumber
\end{align}
while the first and second momenta of the position probability distributions read:
\begin{align}
\label{meanvaluesX}
&X^1 \equiv X= 0 \\ \nonumber
&X^2 \equiv Y= 0 \\ \nonumber
&X^3 \equiv Z= 0 \\ \nonumber
&(X^1)^2= 
\frac{\hbar^2}{p^2_0}
\left( \begin{array}{ccc}
\frac{1}{2}  u_1(a) - u_2(a)\left( \frac{1}{2} + \frac{1}{4 a} \right) + \frac{3}{8a} & 0 & -\frac{u_1(a) + 2 u_2(a)}{4}   \\
0 & 2  u_2(a) -  u_1(a) & 0 \\
-\frac{u_1(a) + 2 u_2(a)}{4}  
& 0 & \frac{1}{2}  u_1(a) - u_2(a)\left( \frac{1}{2} + \frac{1}{4 a} \right) + \frac{3}{8a} \\
\end{array} \right) \\ \nonumber
&(X^2)^2= 
\frac{\hbar^2}{p^2_0}
\left( \begin{array}{ccc}
\frac{1}{2}  u_1(a) - u_2(a)\left( \frac{1}{2} + \frac{1}{4 a} \right) + \frac{3}{8a} & 0 & \frac{u_1(a) + 2 u_2(a)}{4}   \\
0 & 2  u_2(a) -  u_1(a) & 0 \\
\frac{u_1(a) + 2 u_2(a)}{4}  
& 0 & \frac{1}{2}  u_1(a) - u_2(a)\left( \frac{1}{2} + \frac{1}{4 a} \right) + \frac{3}{8a} \\
\end{array} \right) \\ \nonumber
&(X^3)^2= 
\frac{\hbar^2}{p^2_0} 
\left( \begin{array}{ccc}
\frac{1}{4a} + \frac{4a -1}{8a} u_2(a) - u_1(a) & 0 & 0 \\
0 & \frac{3}{4a} - \frac{12 a + 3 }{4a} u_2(a) + 2 u_1(a) & 0 \\
0 & 0 & \frac{1}{4a} + \frac{4a -1}{8a} u_2(a) - u_1(a) \\
\end{array} \right) \notag\\
\end{align}
The matrices associated to the components of momentum and position
observable along the $x$ and $y$ axes differ one from each other, but share the same spectrum.

\begin{figure}[htbp!] 
\center
\includegraphics[scale=0.5]{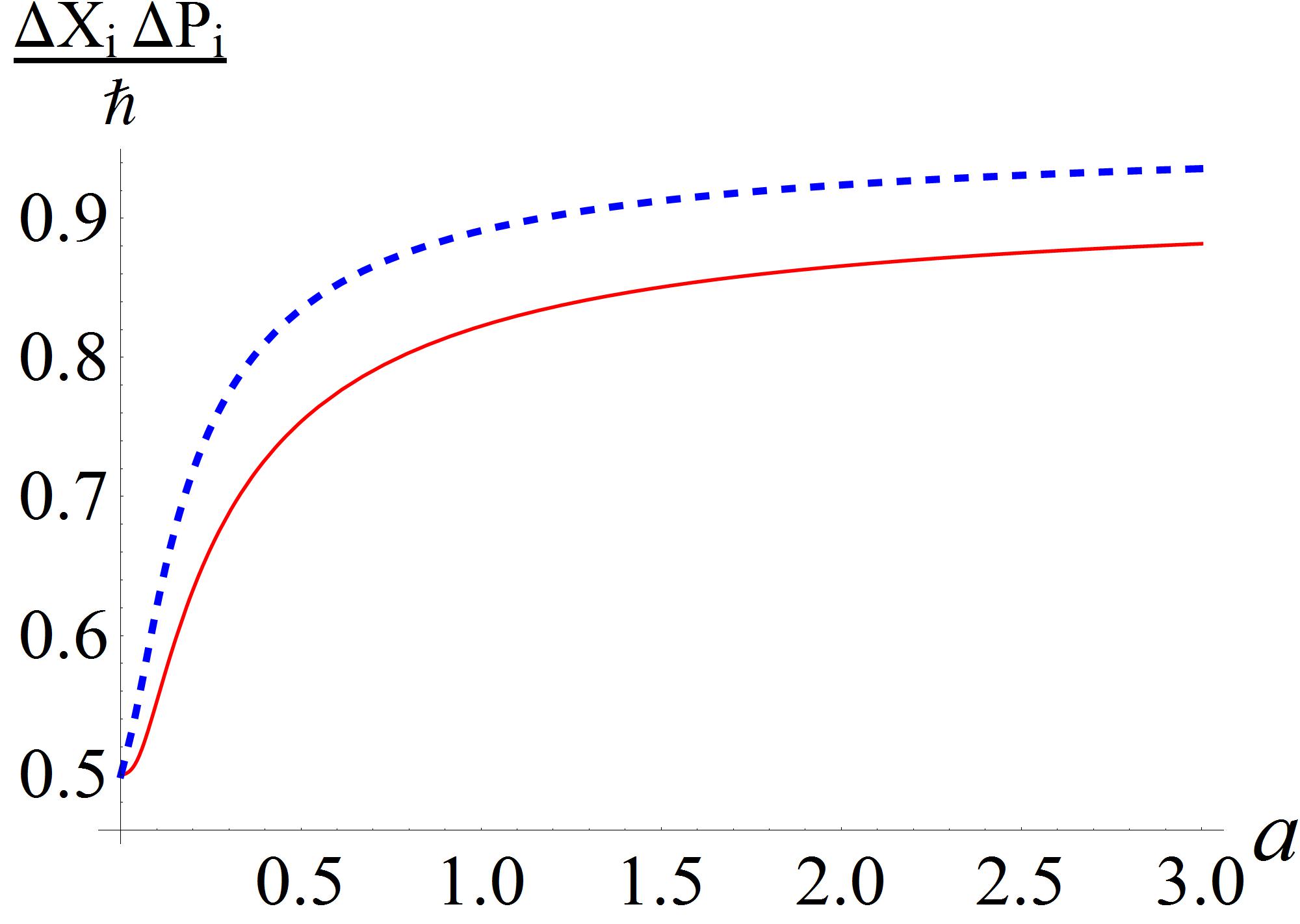}
\caption{(color online): uncertainty products $\Delta X_j \Delta P_j$  in units of $\hbar$, along directions parallel 
(red line) and perpendicular (blue line) to $\vett{p}_0$, for eigenstates of $ S_z $ with eigenvalue $ s = \pm 1$.}
\label{100}
\end{figure}

\begin{figure}[htbp!] 
\center
\includegraphics[scale=0.5]{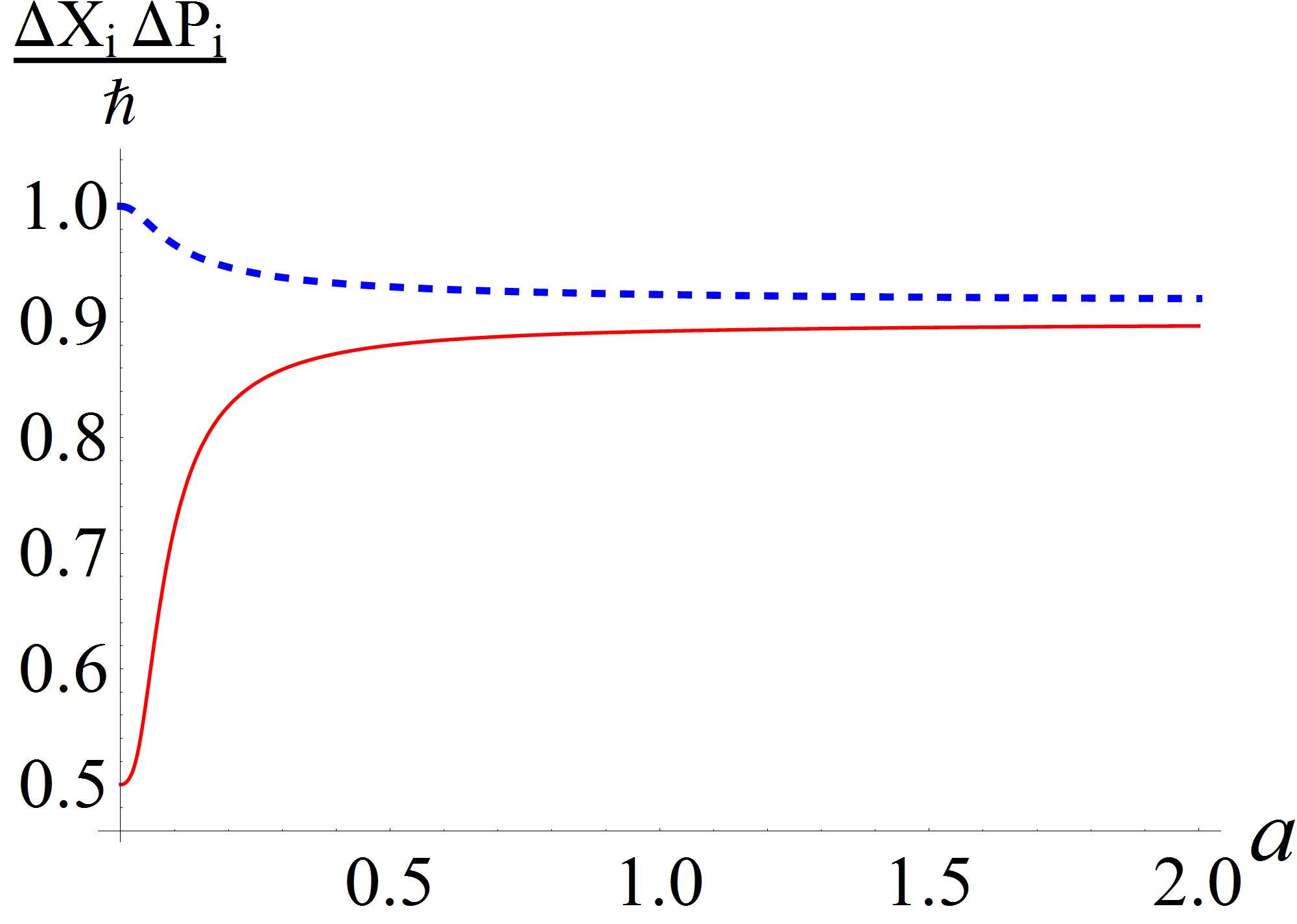}
\caption{(color online): uncertainty products $\Delta X_j \Delta P_j$  in units of $\hbar$, along directions parallel 
(red solid line) and perpendicular (blue dashed line) to $\vett{p}_0$, for eigenstates of $ S_z $ with eigenvalue $ m_s = 0$.}
\label{010}
\end{figure}

The uncertainty products on the $z$ and $x (y)$ axis, which correspond parallel and perpendicular to $\vett{p}_0$, are given by:
\begin{itemize}
\item if $\vett{h}=(1,0,0)$, or $ h=(0,0,1) $:
\begin{align}\label{eq:heis_100}
\Delta Z \cdot \Delta P_z &=\hbar \frac{ \sqrt{ \left[ 1 + 16 a^2 \, u_1(a) - \frac{\left( 1 - 2 a \, u_1(a) \right)^2}{1 - 2 a \, u_2(a)} \right] \Big[ \frac{1}{4a} + \left( \frac{1}{2} - \frac{1}{8a} \right) u_2(a) - u_1(a) \Big] } }{1 - 2 a \, u_2(a)} \\
\Delta X \cdot \Delta P_x &=\hbar \frac{\sqrt{ 2a\left( 1 - 4a\,u_2(a) \right)\left(\frac{1}{2}u_1(a)- u_2(a)\left( \frac{1}{2}+\frac{1}{4a} \right) +\frac{3}{8a}\right)}}{1 - 2a\, u_2(a)}\\
\end{align}
\item if $\vett{h}=(0,1,0)$ 
\begin{align}\label{eq:heis_010}
\Delta Z \cdot \Delta P_z &=\hbar \frac{ \sqrt{ \frac{1}{a}\left[\left( 1-8au_1(a) \right)u_2(a)-u_1^2(a)\right]\left(\frac{3}{4a} - \left( 3 + \frac{3}{4a} \right) u_2(a) + 2 u_1(a)\right) }}{2 u_2(a)} \\
\Delta X \cdot \Delta P_x &=\hbar \frac{\sqrt{u_1(a)\left(2u_2(a)-u_1(
a)\right)}}{u_2(a)}
\end{align}
\end{itemize}
These results are presented in figures \eqref{100} and \eqref{010} respectively.

As highlighted in the case of states with definite polarization, the preparation uncertainty relations explicitly show a highly non-trivial dependence on the width parameter $a$.
For $\textbf{h}=(1,0,0)$ ($=(0,0,1)$) the uncertainty products are monotonically increasing functions of $a$, converging to $\frac{\hbar}{2}$ in the $a \to 0$ limit and to finite
values in the opposite $a \to \infty$ limit.
The asymptotic expansions of \eqref{eq:heis_100} and \eqref{eq:heis_010} are listed, in units of $ \hbar $, in Table \ref{table:limiting2}.

For $\textbf{h} = (0,1,0)$, on the other hand, this behavior is exhibited only by the uncertainty product along the direction 
of $\vett{p}_0$: the other uncertainty product appears a monotonically decreasing function of $a$, converging to $1$ in the $a \to 0$ limit.
The apparently anomalous behaviour of the uncertainty product in figure \ref{010} is a consequence of the fact
that the state \eqref{spindefinite} has, for decreasing $ a $, vanishing projection on the physically relevant space $\hsys$.

$ $
\begin{table}[htbp]
\begin{center}
\begin{tabular}{c c c c}
\vspace{0.001in} \\
\hline\\
$\mathbf{h}$ & axis & small $a$ & large $a$ \tabularnewline
\vspace{0.001in}\\
\hline\\
\vspace{0.001in} \\
$(1,0,0)$ & 
z & $\frac{1}{2} + 4a^2 $ & $\frac{\sqrt{21}}{5} - \frac{59 \sqrt{3}}{350 \sqrt{7} a}$ \\
\vspace{0.001in} \\
$(1,0,0)$ & x & $\frac{1}{2} + a + 40 a^3$ & $\frac{3\sqrt{41}}{2} - \frac{1381}{2800 \sqrt{41} a}$ \\
\vspace{0.001in} \\
$(0,1,0)$ & z & $\frac{1}{2} + 16 a^2 + 272 a^3$ & $\frac{9}{10} - \frac{1}{175 a}$ \\
\vspace{0.001in} \\
$(0,1,0)$ & x & $1-8a^2 + 32 a^3$  & $\frac{\sqrt{21}}{5} + \frac{2 \sqrt{3}}{175 \sqrt{7} a}$ \\
\vspace{0.001in} \\
\hline\\
\end{tabular}
\label{table:limiting2}
\end{center}
\caption{Limiting behaviour of the uncertainty product \eqref{heisenberg}, in units of $ \hbar $, for several choice of the vector $ \mathbf{h} $ (first column), along directions parallel (z axis) and perpendicular (x axis) to $ \vett{p}_0 $ (second column). The expansions relative to the vector $ \vett{h} = (0,0,1) $ are not listed being equal to the case $ \vett{h} = (1,0,0) $.}
\end{table}

Finally, \eqref{eq:heis_100} and \eqref{eq:heis_010} depend quadratically on $ \mathbf{h}\in\C^3 $: in Appendix \eqref{app:max}, the minimum and maximum valued attained by the uncertainty product for different choices of $\mathbf{h}$ have been calculated and plotted over a meaningful range of values for the parameter $a$, and the results compared with those shown in figures \eqref{100}, \eqref{010}.

\subsubsection{Probability Distribution of $S_z$}

The explicit expression of the probability distribution of $S_z$ for a spin state of the form \eqref{spindefinite} will now be given. It reads:
\begin{equation}
\label{psz}
p(S_z= \hbar m_s) = \frac{\sum_{ij}\mathbf{h}_i^* \, [\Sigma (m_s)]_{ij} \, \mathbf{h}_j}{K}
\end{equation}
where $m_s = 2-s$,
\begin{equation}
[\Sigma (m_s)]_{ij} = \int \frac{d^3\vett{p}}{p^0} \, [\Pi(\vett{p})]^*_{i s} \, [\Pi(\vett{p})]_{s j} \, g^2(\vett{p})
\end{equation}
and $ \mathbf{K} $ is the normalization constant introduced above.

Equation \eqref{psz} is then the ratio between two quadratic functions of $\mathbf{h}$. 
The denominator is explicitly given by \eqref{meanvaluesP}, while the matrix $ {\Sigma (m_s)} $ 
in the numerator is most conveniently expressed in terms of the function $u_4(a) = 1-2a u_2(a)$ 
and reads:
\begin{equation}
\begin{split}
{\Sigma (1)} &= \begin{pmatrix}
4a + (1-6a) u_4(a) & 0 & 0 \\
0 & 2 (1+12a) u_4(a)-(1+8a) & 0 \\
0 & 0 & 1+4a - (1+6a) u_4(a) \\
\end{pmatrix} \\
{\Sigma (0)} &= \begin{pmatrix}
4a-(2a+24a^2)u_2(a) & 0 & 0 \\
0 & (1+6a)8u_2(a)-4a & 0 \\
0 & 0 & 4a-(2a+24a^2)u_2(a) \\
\end{pmatrix} \\
{\Sigma (-1)} &= \begin{pmatrix}
1+4a - (1+6a) u_4(a) & 0 & 0 \\
0 & 2 (1+12a)u_4(a)-(1+8a) & 0 \\
0 & 0 & 4a + (1-6a) u_4(a) \\
\end{pmatrix}
\end{split}
\end{equation}
from which it is evident that: \\
a) $p(S_z=1) = p(S_z=-1)$ for the spin state relative to the choice $\mathbf{h}=(0,1,0)$ \\ 
b) $p(S_z=1)$ for the spin state with $\mathbf{h}=(1,0,0)$ and $p(S_z=-1)$ for the spin state with $\mathbf{h}=(0,0,1)$ are equal to each other.

\begin{figure}[!htbp]
{\bf (a)}
\\
\center
\includegraphics[scale=0.6]{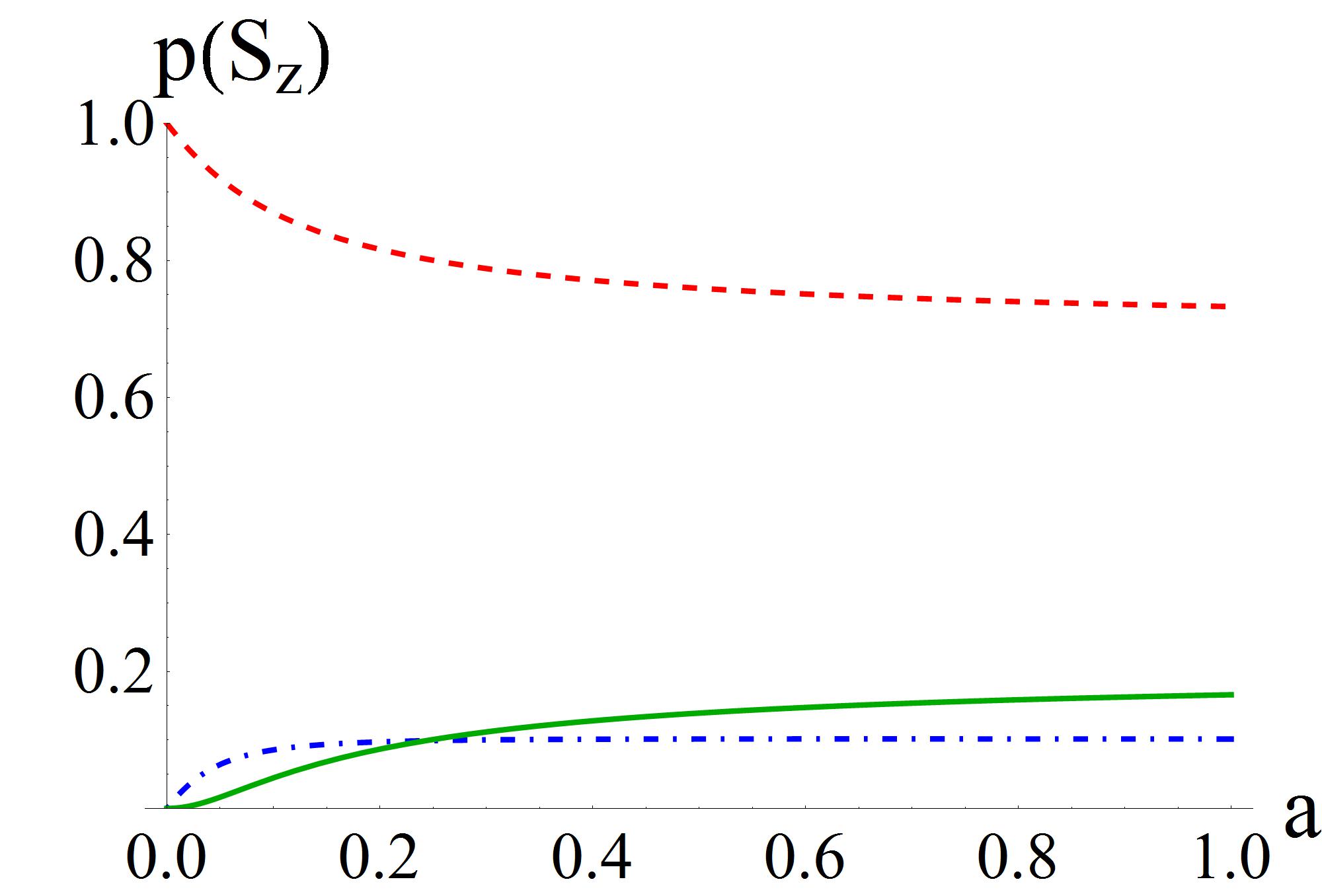}
\vspace{.5truecm}
\\
{\bf (b)}
\\
\center
\includegraphics[scale=0.6]{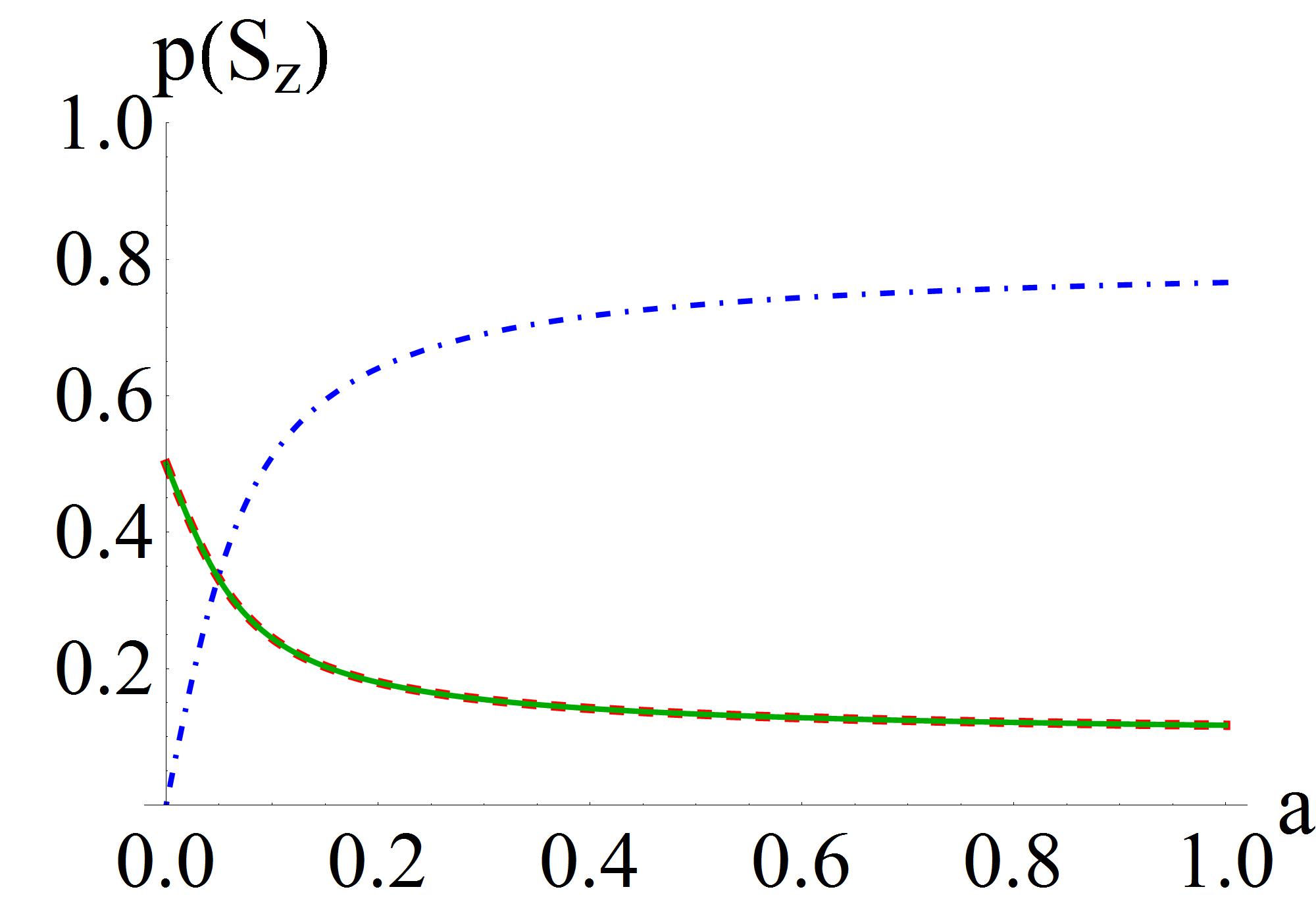}
\vspace{.5truecm}
\\
{\bf (c)}
\\
\center
\includegraphics[scale=0.6]{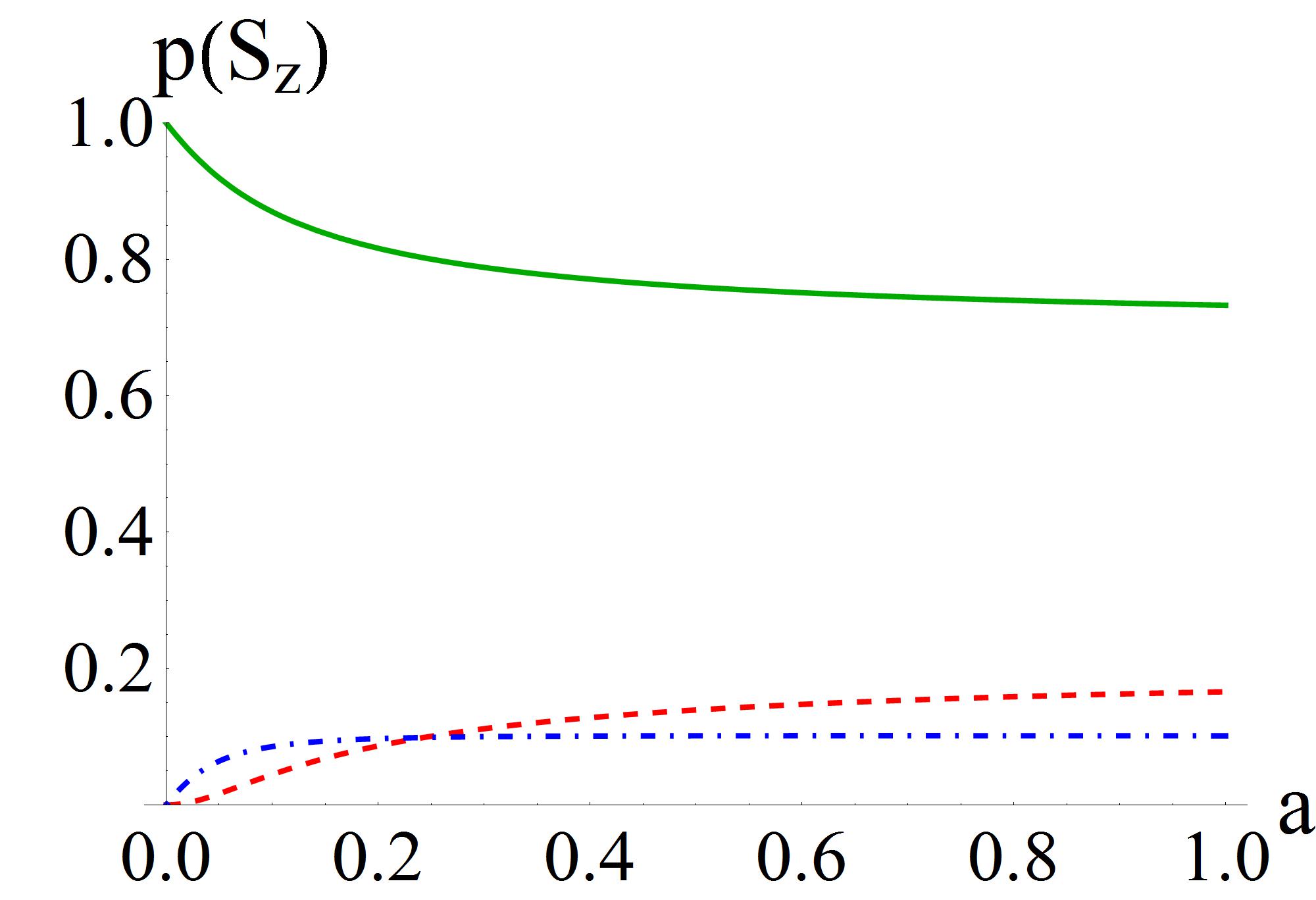}
\caption{Color online: probability distribution of $ S_z $ (red dashed line for the eigenvalue $ m_s=1 $, green solid line for $ m_s=0 $, blue dot-dashed line for $ m_s=-1 $) on the spin states \eqref{spindefinite} with $\mathbf{h} = \mathbf{e}_1$ (a), $\mathbf{e}_2$ (b) and $\mathbf{e}_3$ (c)}
\label{fig:spinddpe1}
\end{figure}

Figure \ref{fig:spinddpe1} shows the cases $\mathbf{h} = \mathbf{e}_i$.

The asymptotic expansions of the probability distributions for the spin $ S_z $ are listed in Table \eqref{table:sz}.

\begin{table}[htbp!]
\begin{center}
\begin{tabular}{c c c c}
\hline \\
$S_z/\hbar$ & $\mathbf{h}$ & small $a$ & large $a$ \\
\\
\hline\\
 1 & (1,0,0) &    $1-2a+8a^2$ & $\frac{7}{10}+\frac{51}{1400 a}$ \\ 
 0 & (1,0,0) &      $2a-16a^2$ & $\frac{1}{10}+\frac{3}{1400 a}$ \\
-1 & (1,0,0) &             $8a^2$ & $\frac{2}{10}-\frac{54}{1400 a}$ \\
 1 & (0,1,0) & $\frac{1}{2}-4a+8a^2$ & $\frac{1}{10} + \frac{3}{175 a}$ \\ 
 0 & (0,1,0) &      $8a-16a^2$ & $\frac{8}{10} - \frac{6}{175 a}$ \\
\hline
\label{table:sz}
\end{tabular}
\end{center}
\caption{Asymptotic behaviour of the probability distribution of $S_z$ for the choices of $\mathbf{h}$ reported in figure \ref{fig:spinddpe1}.}
\end{table}

Due to the projection onto the physical Hilbert space, the spin of the photon therefore ceases to be a definite quantity and is characterized by a probability distribution resulting by integration \eqref{pp} over the momentum variable.
This result is consequence of the fact that states with definite spin exist only in the extended Hilbert space and puts under an even more clear evidence the enhancement of the statistical character of the quantum theory stemming from the unsharpness of POVMs.

\section{Conclusions}
\label{sec:Conclusions}

In the present work, we have constructed a unified procedure for treating all single-photon
observables as POVMs, including the notoriously delicate position and spin observables, and
for computing all joint probability distributions.

Starting from the representation theory of the Poincar\'e group for spin $s=1$ and mass $m=0$ particles,
we have demonstrated how the suppression of the $0$-helicity component of the single-photon
wavefunction, which corresponds through the isomorphism $ V $ to the suppression of the 
longitudinal component, can be viewed as the action of a projection operator from an extended 
Hilbert space onto the single-photon Hilbert space.

We have then shown how this theoretical construction naturally brings along the notion of POVMs, 
since \textit{any} observable can be introduced as a PVM on the extended Hilbert space, the form
of which is very naturally suggested by the well-established quantum theory of relativistic 
particles with mass $m>0$ and spin $s=1$, and then turned, by means of the projection, into a 
directly manageable POVM on $\hsys$.

The suppression of $0$-helicity states determines a coupling between momentum and spin,
expressed by the sharpness of the helicity observable. The intrinsic unsharpness which 
characterizes the POVMs of spin and position, on the other hand, reflects the impossibility of
preparing a photon with definite spin along a fixed arbitrary direction or localizing it in a
bounded region of the configuration space, according to the Paley-Wiener theorem.

We have explicitly derived the joint probability distributions for spin and momentum, and for
spin and position, along with their marginals and with the probability distribution of helicity.

Finally, we have extensively investigated the consequences of the introduction of the POVMs, 
i.e. the effects of the projection $\YUPPI$, by studying the preparation uncertainty relations 
for position and momentum and the probability distribution of spin $ S_z $. 
We have considered different classes of Gaussian states, which in the non-relativistic context saturate 
the inequality $\Delta X_k \Delta P_k \geq \frac{\hbar}{2}$, namely those with definite polarization 
and with definite spin along a spatial axis.
The results we have found show a clear enhancement of the statistical character of position and 
spin observables brought into stage by the appearance of POVMs, which ranges from the increment 
of the uncertainty products for any value of the Gaussian width $a$, to the impossibility of 
preparing a state with definite spin along an axis independent on momentum.
Only in the limiting case of infinitely sharp wavefunction in the momentum space (which corresponds 
to $ a \to 0 $ ), a single-photon state can have a defined value of the spin and the uncertainty
products return to the value $ \frac{\hbar}{2}$.
Such behaviour is observed both for Gaussian states with definite polarization and spin, confirming 
the coherence of the underlying formalism, which can be immediately applied to all single-photon 
observables, and to all single-photon states and density matrices.

We believe there exist favourable prospects for this formalism to be applied in a high-precision description of interference phenomena at the basis of experiments with single photons \cite{Sciarrino,Bomba,Larque,Beveratos}. 

\section*{Acknowledgements}

We express our deep gratitude to Dr. Andrea Smirne for support and fruitful discussions in merit, and to Prof. Luca Molinari and Dr. Davide Galli for their useful feedback.
M.M. also acknowledges fundings from the Dr. Davide Colosimo Award, celebrating the memory of physicist Davide Colosimo.

\appendix

\section{Maximization and Minimization of $\Delta X_k \Delta P_k$ for Spin States}
\label{app:max}

In this Appendix we identify the quantities $ \mathbf{h}_{max} $ and $ \mathbf{h}_{min} $ which respectively maximize and minimize the Heisenberg Uncertainty Relations \eqref{eq:heis_100} and \eqref{eq:heis_010}.

\begin{figure}[htbp!]
{\bf (a)}
\\
\center
\includegraphics[scale=0.55]{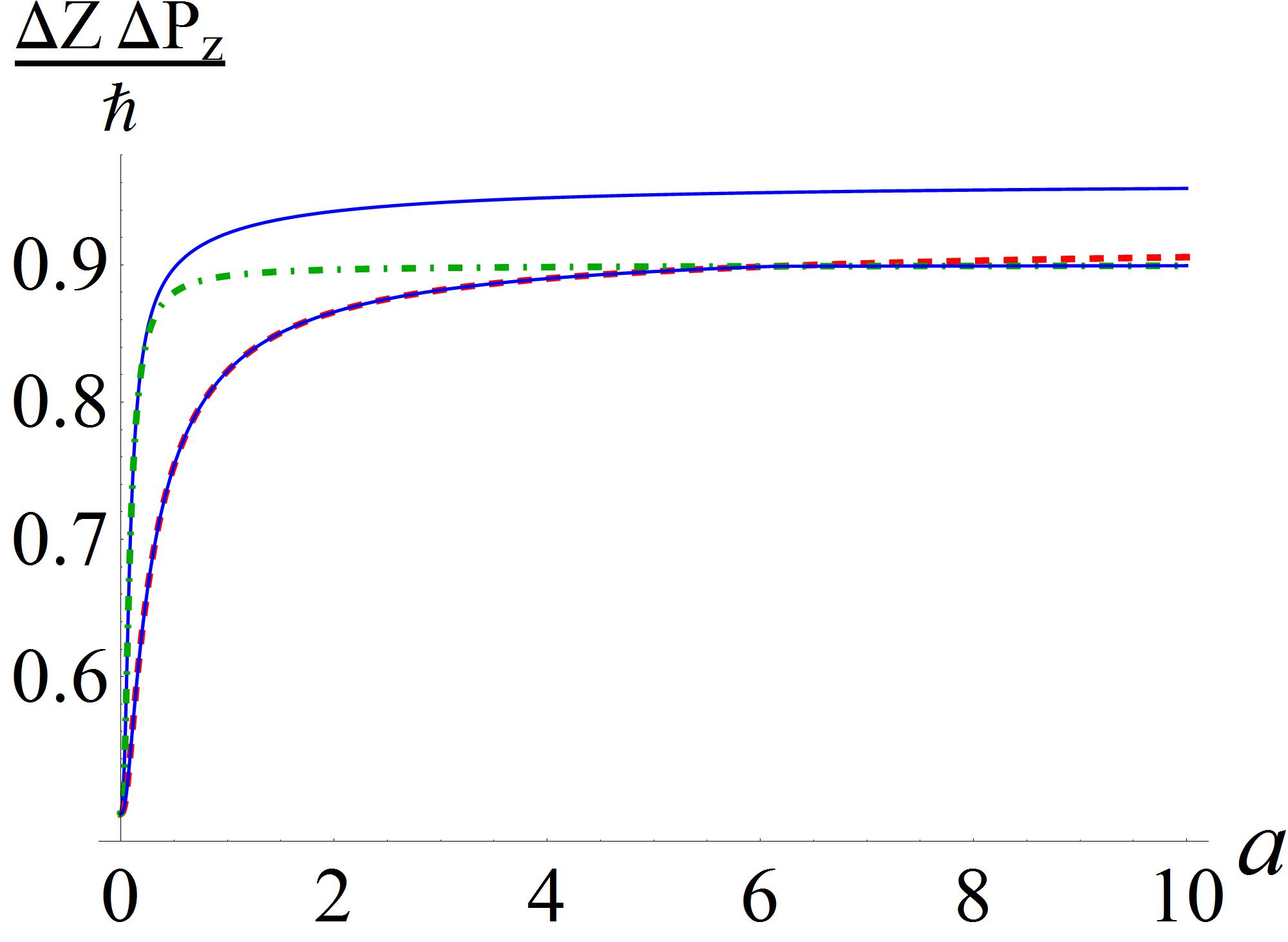}
\vspace{.5truecm}
\\
{\bf (b)}
\\
\center
\includegraphics[scale=0.55]{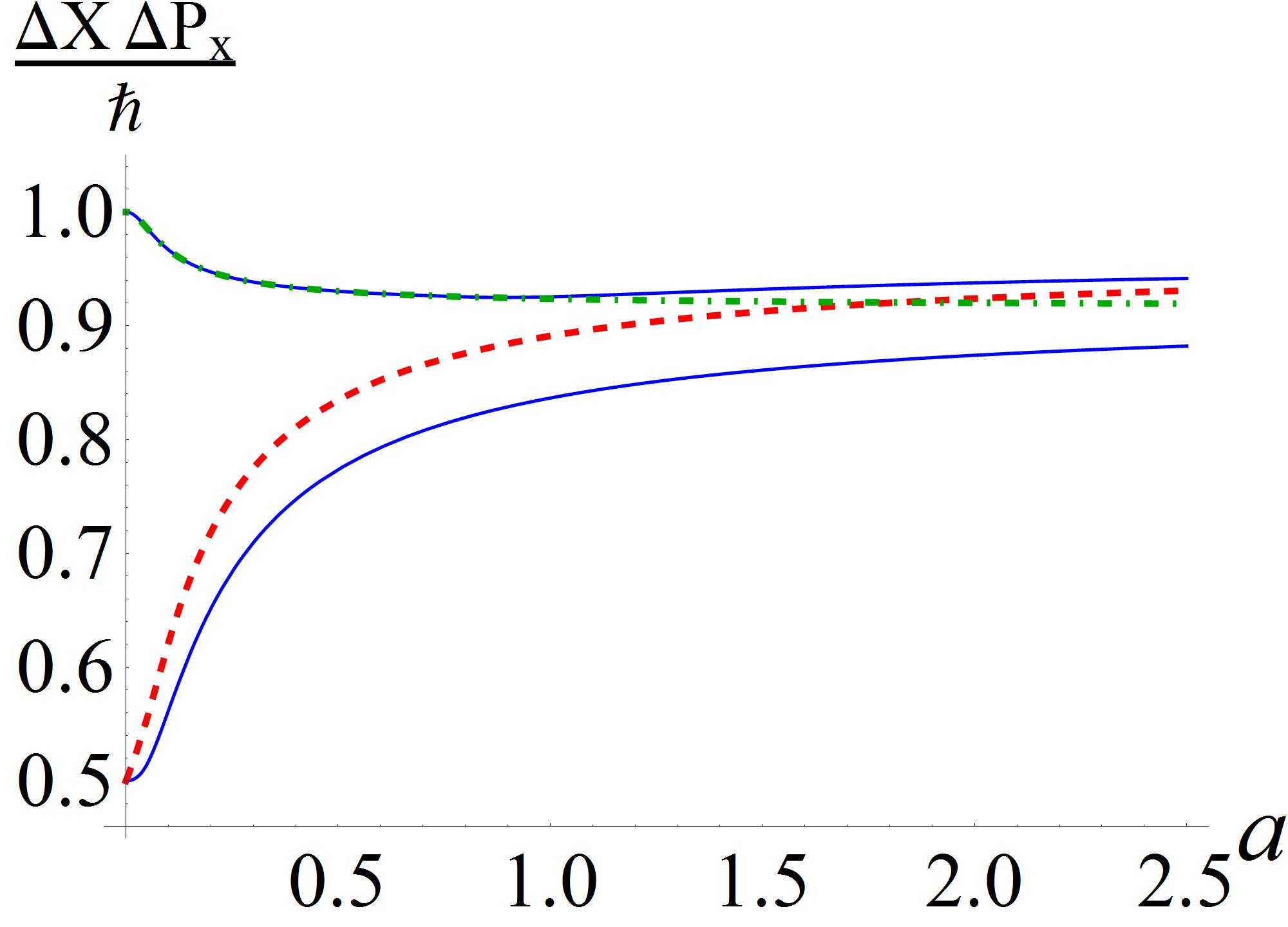}
\caption{(color online) minimum and maximum (blue lines) of the Heisenberg uncertainty product along directions parallel (left panel) and perpendicular (right panel) to $\vett{p}_0$, in comparison with $\mathbf{h}=(1,0,0)$ (red dotted line) and $\mathbf{h}=(0,1,0)$ (green dot-dashed line).}
\label{fig:h1h2h3_guarmotta} 
\end{figure}

For the purpose of simplifying the forthcoming calculations, let us parametrize $\mathbf{h} \in \mathbb{C}^3$ as follows:
\begin{equation}
\mathbf{h} = \left(
\begin{array}{c}
\frac{\tilde{h}_1}{\sqrt{1 - 2 a u_1(a)}} \\
\frac{\tilde{h}_2}{\sqrt{4 a u_2(a)}} \\
\frac{\tilde{h}_3}{\sqrt{1 - 2 a u_1(a)}} \\
\end{array}
\right)
\end{equation}
It can be easily verified that $\mathbf{h} \cdot K \mathbf{h}  = 1$ if and only if $|\tilde{h}_1|^2 + |\tilde{h}_2|^2 + |\tilde{h}_3|^2 = 1$. This choice leads to the following expressions:
\begin{equation}
\begin{split}
\label{av_z_h1h2h3}
&\braket{P_z} = p_0 \, \left[ \frac{1 - 2 a u_1(a)}{1 - 2 a u_2(a)} (1-\rho) + \frac{u_1(a) }{u_2(a)} \rho \right] \\
\quad\quad\quad
&\braket{(P_z)^2} = p_0^2 \, \left[ \frac{1 + 16 a^2 u_1(a)}{1 - 2 a u_2(a)} (1-\rho) + \frac{1 - 8 a u_1(a)}{u_2(a)} \rho \right] \\
\\
&\braket{Z} = 0 \\
\quad\quad\quad
&\braket{Z^2} = \frac{\hbar^2}{p_0^2} \, \, \left[ \frac{\frac{1}{4a} + \frac{u_2(a)}{2} + \frac{u_2(a)}{8a} - u_1(a)}{1 - 2 a u_2(a)} (1-\rho) + \frac{\frac{3}{4a} - \left( 3 + \frac{3}{4a} \right) u_2(a) + 2 u_1(a)}{4 a u_2(a)} \rho \right] 
\end{split}
\end{equation}
where $|\tilde{h}_2|^2 = \rho$, $|\tilde{h}_1|^2 + |\tilde{h}_3|^2 = 1-\rho$. The square of the Heisenberg product
constructed from \eqref{av_z_h1h2h3} is a polynomial of third degree in the variable $\rho$ possessing, for each fixed
value of $a$, a global maximum at a value $\rho_{max}(a)$ in the interval $(0,1)$ and a global minimum at
$\rho_{min}(a) = \Theta(a-a^*)$, where $a^* \sim 6.13116$.
We do not detail the rather intricate analytic form of the minimum and maximum values of the Heisenberg product, but 
limit ourselves to show, in figure \ref{fig:h1h2h3_guarmotta}, their values against those 
in figures \ref{100} and \ref{010} relative to the choices $\mathbf{h}=(1,0,0)$ and 
$\mathbf{h}=(0,1,0)$.

On the $ x $ axis, it is convenient to parametrize also $\tilde{h}$ as follows:
\begin{equation}
\begin{split}
\tilde{\mathbf{h}} = \left(
\begin{array}{c}
\sqrt{\lambda} \, e^{i \phi_1}\\
\sqrt{1-\lambda} \sqrt{\xi} \\
\sqrt{1-\lambda} \sqrt{1-\xi} \, e^{i \phi_2}\\
\end{array} \right)
\end{split}
\end{equation}
obtaining:
\begin{equation}
\begin{split}
\label{av_x_h1h3}
&\braket{\hat{P}_x} = - (2 p_0) \, \frac{2 a u_1(a)}{\sqrt{\left( 1 - 2 a u_2(a) \right) \left( 4 a u_2(a) \right) }} \, \, (1-\lambda) \, \sqrt{\xi-\xi^2} \, \cos(\phi_2)
\\
\notag\\
&\braket{(\hat{P}_x)^2} = (2 p_0)^2 \, a \, \left[ \frac{1 - 2 a u_1(a)}{2 (1 - 2 a u_2(a))} \lambda + \frac{1 -6 a u_1(a)}{2 (1 - 2 a u_2(a))} (1-\lambda) (1-\xi) + \frac{u_1(a)}{u_2(a)} (1-\lambda) \xi \right] \\
\notag\\
&\braket{\hat{X}} = 0 \\
 \notag\\
&\braket{\hat{X}^2} = \frac{\hbar^2}{(2 p_0)^2} \, \Bigg[ \frac{ u_1(a) - \frac{u_2(a)}{a} - 4 u_2(a) + \frac{3}{2a}}{1 - 2 a u_2(a)} \, \lambda + \notag\\
&+ \frac{ 3 \, u_1(a) - u_2(a) + \frac{3}{2a}}{1 - 2 a u_2(a)} \, (1-\lambda) (1-\xi) + \frac{2u_2(a)-u_1(a)}{a u_2(a)} \, (1-\lambda) \xi \Bigg] \\
\end{split}
\end{equation}

The square of the Heisenberg uncertainty product is a linear and monotonically decreasing function of $\cos^2(\phi_2)$.
The minimum is attained at $\cos^2(\phi_2)=1$, $\xi=0$ and $\lambda = \Theta(a^*-a)$, $a^* \sim 2.6095$, and the
maximum at $\cos^2(\phi_2)=0$, $\xi = \xi_{max}(a)$ and $\lambda=0$. $\xi_{max}(a)$ results from a straightforward 
but quite lengthy maximization procedure. 
The minimum and maximum values of the Heisenberg uncertainty product are shown in Fig. \ref{fig:h1h2h3_guarmotta} against the corresponding values \eqref{eq:heis_100} for $\mathbf{h}=(1,0,0)$ and $\mathbf{h}=(0,1,0)$.

Finally the asymptotic expansions of the Heisenberg Uncertainty Relations for small and large values of the parameter $ a $ are listed in Table \ref{table:limiting3} in units of $ \hbar $.

$ $
\begin{table}[htbp]
\begin{center}
\begin{tabular}{c c c c}
\vspace{0.001in} \\
\hline\\
$\mathbf{h}$ & axis & small $a$ & large $a$ \tabularnewline
\vspace{0.001in}\\
\hline\\
\vspace{0.001in} \\
$ \mathbf{h}_{min}(a) $ & z & $\frac{1}{2}+4a^2 $  & $\frac{9}{10} - \frac{1}{175 a}$ \\
\vspace{0.001in} \\
$ \mathbf{h}_{max}(a) $ & z & $\frac{1}{2}+\frac{a}{2}+12a^2$  & $2\sqrt{\frac{3}{13}} - \frac{317}{3640 a}\sqrt{\frac{3}{13}} $ \\
\vspace{0.001in} \\
$ \mathbf{h}_{min}(a) $ & x & $\frac{1}{2}+4a^2$  & $\frac{9}{10} - \frac{23}{525 a}$ \\
\vspace{0.001in} \\
$ \mathbf{h}_{max}(a) $ & x & $1 - 8a^2 + 32 a^3$  & $2\sqrt{\frac{3}{13}} - \frac{1021}{9100} \sqrt{\frac{3}{13}} \, \frac{1}{a} $ \\
\vspace{0.001in} \\
\hline\\
\end{tabular}
\label{table:limiting3}
\end{center}
\caption{Limiting behaviour of the Heisenberg Uncertainty Relations \eqref{heisenberg}, in units of $ \hbar $, for $ \mathbf{h}_{min} $ and $ \mathbf{h}_{max} $ (first column), along directions parallel (z axis) and perpendicular (x axis) to $ \vett{p}_0 $ (second column).}
\end{table}

\bibliographystyle{ieeetr}

\begin{thebibliography}{99}
\bibitem{tsang} M. Tsang, {\em{Phys. Rev. Lett.}} {\bf{102}}, 253601 (2009)
\bibitem{Sciarrino} G. Vallone, V. D'Ambrosio, A. Sponselli, S. Slussarenko, L. Marrucci, F. Sciarrino
and P. Villoresi, {\em{Phys. Rev. Lett.}} \textbf{113}, 060503 (2014)
\bibitem{Birula} I. Bialynicki-Birula, {\em{Phys. Rev. Lett.}} \textbf{80}, 5247 (1998) 
\bibitem{Chiao} R. Y. Chiao and Y.-S. Wu, {\em{Phys. Rev. Lett.}} \textbf{57}, 933 (1986)
\bibitem{Tomita} A. Tomita and R. Y. Chiao, {\em{Phys. Rev. Lett.}} \textbf{57}, 937 (1986)
\bibitem{Bomba} G.S. Paraoanu , {\em{Phys. Rev. Lett.}} \textbf{97}, 180406 (2006)
\bibitem{Chen} Y. Liu et. al., {\em{Phys. Rev. Lett.}} {\bf{111}}, 130502 (2013)
\bibitem{Grier} D. G. Grier, Nature \textbf{424}, 810-816 (2003).
\bibitem{newton} T.D. Newton and E.P. Wigner, {\em{Rev. Mod. Phys.}} {\bf{21}}, 400 (1949)
\bibitem{wightman} A.S. Wightman, {\em{Rev. Mod. Phys.}} {\bf{34}}, 845 (1962)
\bibitem{KrausPos}  K. Kraus "Position Observable for the Photon" in "Uncertainty Principles and the Foundation of Quantum Mechanics" W.C. Price, S.S. Chissich, ed. Wiley (1977)
\bibitem{Holevo} A. Holevo, "Statistical structure of quantum theory", Lecture Notes in Physics Monographs, Springer (2001)
\bibitem{HolevoPOVM} A.S. Holevo, "Probabilistic and statistical aspects of quantum theory", North-Holland Publ. Cy., 
Amsterdam (1982).
\bibitem{Holevo2} A. S. Holevo, {\em{Rep. Math. Phys.}} {\bf{13}}, 3 (1978)
\bibitem{Holevo3} A. S. Holevo, {\em{Russian Microelectronics}} {\bf{13}}, 21 (2006)
\bibitem{mandel} L. Mandel and E. Wolf, Optical Coherence and Quantum Optics (Cambridge University Press, Cambridge, 1995).
\bibitem{HawtonPOVM} M. Hawton {\em{Phys. Rev. A}} {\bf{82}}, 012117 (2010)
\bibitem{sp1} A. I. Akhiezer and V. B. Berestetskii, Quantum electrodynamics
(Interscience Publishers, New York, 1965).
\bibitem{sp2} S. J. van Enk and G. Nienhuis, {\em{Europhys. Lett.}} {\bf{25}}, 497 (1994), {\em{J. Mod. Opt.}} {\bf{41}}, 963 (1994).
\bibitem{sp4} K. Y. Bliokh, M. A. Alonso, E. A. Ostrovskaya and A. Aiello, {\em{Phys. Rev. A}} {\bf{82}}, 063825 (2010)
\bibitem{Ludwig} G. Ludwig, An axiomatic Basis for Quantum Mechanics, Berlin: Springer(1985)
\bibitem{Moses} H.E. Moses, {\em{J. Math. Phys.}} \textbf{8}, 1134 (1967);
                H.E. Moses, {\em{J. Math. Phys.}} \textbf{9}, 16 (1968)
\bibitem{Wigner} E. P. Wigner, Ann. Math. \textbf{40}, 149 (1939)
\bibitem{Lahti} P. Busch, M. Grabowski and P. Lahti, Operational Quantum Physics, Lecture Notes in Physics Monographs, Springer (1995)
\bibitem{Breuer2002}
H.-P. Breuer and F.~Petruccione, \emph{The Theory of Open Quantum Systems}
  (Oxford University Press, Oxford, 2002).
\bibitem{Busch2010} P. Busch and G. Jaeger, {\em{Found. Phys.}} {\bf{40}}, 1341 (2010)
\bibitem{Busch1996} P. Busch and A. Shimony, {\em{Stud. Hist. Phil. Mod. Phys.}}
{\bf{27}}, 397 (1996)
\bibitem{Busch1989} P. Busch and F.E. Schroeck Jr. {\em{Found. Phys.}} {\bf{19}}, 807 (1989)
\bibitem{Massar} S. Massar, Phys. Rev. A \textbf{76}, 042114 (2007)
\bibitem{paley_wiener} R. E. A. C. Paley and N. Wiener, Fourier Transforms in the Complex Domain (American Mathematical Society, New York, 1934).
\bibitem{deb2} W.O. Amrein, Helv. Phys. Acta 42 (1969)
\bibitem{Hamermesh} M. Hamermesh, Group theory and its application to physical problems, Dover Publications (1989).
\bibitem{McCabe} G. McCabe, The Structure and Interpretation of the Standard Model, Elsevier (2007)
\bibitem{Mackey} G. W. Mackey, Mathematical Foundation of Quantum Mechanics, Dover (1963)
\bibitem{Levy} J. M. L\'evy-Leblond, {\em{Commun. Math. Phys.}} {\bf{6}}, 286 (1967)
\bibitem{footnoteV} Considering the particular $ s=1 $ unitary and irreducible representation of the spin operators $ \vett{S} $ on $ \C^3 $ and the complexification of $ \R^3 $, one has that the operators $ \vett{S} $ and $ i\hbar\vett{A} $, with $ \vett{A} $ denoting the generators of the rotations in $ \R^3 $, satisfy the same commutation relations. Then, the irreducibility of the $ s=1 $ representation implies that there exists a unitary transformation $ V $ such that $ \vett{S} = V (i\hbar\vett{A}) V^\dagger $.
\bibitem{footnote} The term pseudo-Hilbert is inherited from the usual terminology which refers to the
pseudo-Euclidean character of the relativistic four-dimensional metric.
\bibitem{riemann_silb} I. Bialynicki-Birula and Z. Bialynicka-Birula, J. Phys. A, \textbf{46} 053001 (2013)
\bibitem{Gupta} S. Gupta, {\em{Proc. Phys. Soc.}} \textbf{A63}, 681 (1950)
\bibitem{Bleuler} K. Bleuler, {\em{Helv. Phys. Acta}} \textbf{5}, 567 (1950)
\bibitem{Segal} V. Bach, J. Fr\"{o}hlich and I. M. Sigal, Adv. Math. 137, 299-395 (1998)
\bibitem{weinberg} S. Weinberg, The Quantum Theory of Fields, Cap. (2.5) and (5.9) Cambridge University Press (2000)
\bibitem{Beltrametti} E.G. Beltrametti and G. Cassinelli, The logic of quantum mechanics, Addison-Wesley (1981)
\bibitem{Moretti} V. Moretti, Spectral Theory and Quantum Mechanics, Springer UNITEXT (2013)
\bibitem{Kraus1983} K. Kraus, States, Effects and Operations: Fundamental Notions of Quantum Theory, Springer, Berlin (1983).
\bibitem{wf_birula} I. Bialynicki-Birula, {\em{Acta Physica Polonica}}, {\bf{86}}, 97 (1994), I. Bialynicki-Birula, in \textit{Progress in Optics}, edited by
E. Wolf (Elsevier, Amsterdam, 1996), Vol. XXXVI.
\bibitem{wf_sipe} J. E. Sipe, Phys. Rev. A, \textbf{52},  1875 (1995)
\bibitem{wf_adlard} C. Adlard, E. R. Pike, S. Sarkar, {\em{Phys. Rev. Lett. }} {\bf{79}}, 1585 (1997)
\bibitem{wf_chan} K. W. Chan, C. K. Law, and J. H. Eberly, {\em{Phys. Rev. Lett.}} {\bf{88}}, 100402 (2002)
\bibitem{Busch2009} P. Busch, {\em{Found. Phys.}} {\bf{39}}, 712 (2009)
\bibitem{Abra} M. Abramowitz and I. Stegun Handbook of Mathematical Functions, Dover (1964)
\bibitem{Larque} M. Larqu\'{e}, A. Beveratos, I. Robert-Philip, Eur. Phys. J. D \textbf{47}, 119–125 (2008)
\bibitem{Beveratos} A. Beveratos, R. Brouri, T. Gacoin, A. Villing, J.-P. Poizat, P. Grangier, Phys. Rev. Lett. \textbf{89}, 187901 (2002)


\end{thebibliography}

\end{document}